\documentclass[10pt]{iopart}

\usepackage{iopams}  
\usepackage{units}
\usepackage{textcomp}
\usepackage{amssymb}
\usepackage{stmaryrd}
\usepackage{graphicx}

\usepackage{caption}
\usepackage{subcaption}
\usepackage{multirow}

\begin{document}

\title[Nanofibre-based trap for Rb$_2$ molecule]{Nanofibre-based trap for Rb$_2$ molecule}

\author{M. M{\`a}rquez-Mijares$^{1}$, B. Lepetit$^{2}$, E. Brion$^{2}$}
\address{
$^{1}$Instituto Superior de Tecnolog{\'i}as y Ciencias Aplicadas (InSTEC), Universidad de La Habana, Ave. Salvador Allende 1110, Plaza de la Revoluci{\'o}n, La Habana 10400, Cuba\\
$^{2}$Laboratoire Collisions Agrégats Réactivité, UMR5589, Universit\'e Toulouse III Paul Sabatier, CNRS, F-31062 Toulouse Cedex 09, France}

\ead{brion@irsamc.ups-tlse.fr}


\begin{abstract}
We describe a theoretical proposal of a nanofibre-based trap for a Rb$_2$ molecule prepared in the metastable state $(1)^3\Sigma^+_u$. The 
trapping potential results from the combination of a travelling and a standing-wave fields, both carried by the fundamental guided mode HE$_{11}$ of the fibre. We show that, with an experimentally realistic choice of \textsc{laser} frequencies and powers, one can implement a $\approx 200$ $\mu$K-deep well at $\approx 140$ nm from the fibre surface accomodating for $\approx 500$ translational molecular states.
\end{abstract}

%

\submitto{\JPB}
%
%

\section{Introduction}

For the past twenty years, an important effort has been devoted to the development of quantum light-matter interfaces which are instrumental for future quantum networks \cite{HSP10}. In 2002, K. Hakuta and coauthors first suggested that a stretched optical fibre, whose radius at its waist is less
than the wavelength of the guided light, could be a valuable candidate for such a platform. The strong evanescent component of the guided field can indeed couple to neighbouring particles \cite{PLH02} and be used, e.g., to optically trap single cold atoms in the close vicinity of the fibre \cite{VRS10}. The storage \cite{GMN15,SCA15} and Bragg reflection  \cite{SBK16,CGC16} of guided photons could also be demonstrated in arrays of such trapped atoms along a silica nanofibre. Optical nanofibres moreover appear as promising and versatile setups for the investigation of new non-linear quantum optical effects involving cold atoms, and for applications in quantum computation, communication and simulation \cite{NGN16}. 

Though less extensively explored than atoms, molecular systems have, however, started to be considered for their potential applications in quantum technologies \cite{WFF20}. 
It has long been known that a qubit of information can be typically stored in two low-lying long-lived energy eigenstates of a molecule \cite{DeM02, LO05, YKC06, OZR11, BCG18}, while alternative approaches resort to vibrational or spin degrees of freedom \cite{TdV02, BCC15}. Recently, a theoretical scheme was put forward to robustly encode quantum information in the rotational states of individual molecules \cite{ACP20}.
In view of their rich internal structure, ultracold polar molecules trapped in arrays of optical tweezers were also reckoned as a promising tool for quantum simulation of many-body physics \cite{MBZ06} and robust storage and transmission of quantum information \cite{BAB21, NRG18, GBB21}. The parallel preparation of five NaCs molecules was recently achieved in such a structure which offers full internal and motional state control \cite{ZPC22}.

In the present work, we take the first step towards building a platform for quantum technologies which combines the potentialities of nanofibres and \emph{ free} single molecules -- we note that an optical-nanofibre-based interface was already implemented for single organic molecules embedded within a crystal \cite{SPS18}. More precisely, our goal here is to theoretically investigate how to trap the diatomic molecule Rb$_2$ in the vicinity of a silica optical nanofibre.
The electronic structure of the Rb$_2$ molecule has been the subject of several theoretical studies \cite{BRC11,AA12}, stimulated by the possibility to photoassociate and manipulate cold molecules. The ground state $(1)^1\Sigma^+_g$ and the metastable state $(1)^3\Sigma^+_u$ both correlate in the asymptotic limit to a pair of atoms in their ground state, $5s\,^2S_\frac12$. In particular, the metastable state $(1)^3\Sigma^+_u$ was experimentally produced in a free-space photoassociation scheme involving the excited state $^3\Pi_g$ which correlates to the asymptotic limit $5s\,^2S_\frac12$+$5p\,^2P_\frac12$ \cite{BRC11}. It was also suggested that the photoassociation process may be made more efficient through coupling to a photonic crystal \cite{PKH17}.
Here, we show that the Rb$_2$ molecule prepared in its metastable state $(1)^3\Sigma^+_u$ can be optically trapped in the vicinity of an optical nanofibre by the combination of a travelling and a standing-wave fields  both carried by its fundamental guided mode HE$_{11}$. 

Our article is structured as follows. In section \ref{System}, we present the system, recall basic equations and provide useful molecular data. In section \ref{Results}, we present our numerical results, provide the frequencies and intensities of the trapping fields as well as the shape and depth of the resulting trapping potential. We also discuss the interest and limitations of our results -- in particular we justify why the Casimir-Polder force acting on the molecule may be safely neglected. Finally, we conclude in section \ref{Conclusion}, and give perspectives of our work. Complementary information is given in appendices.

\section{Presentation of the system\label{System}} 
\begin{figure*}
\begin{centering}
\includegraphics[width=14cm]{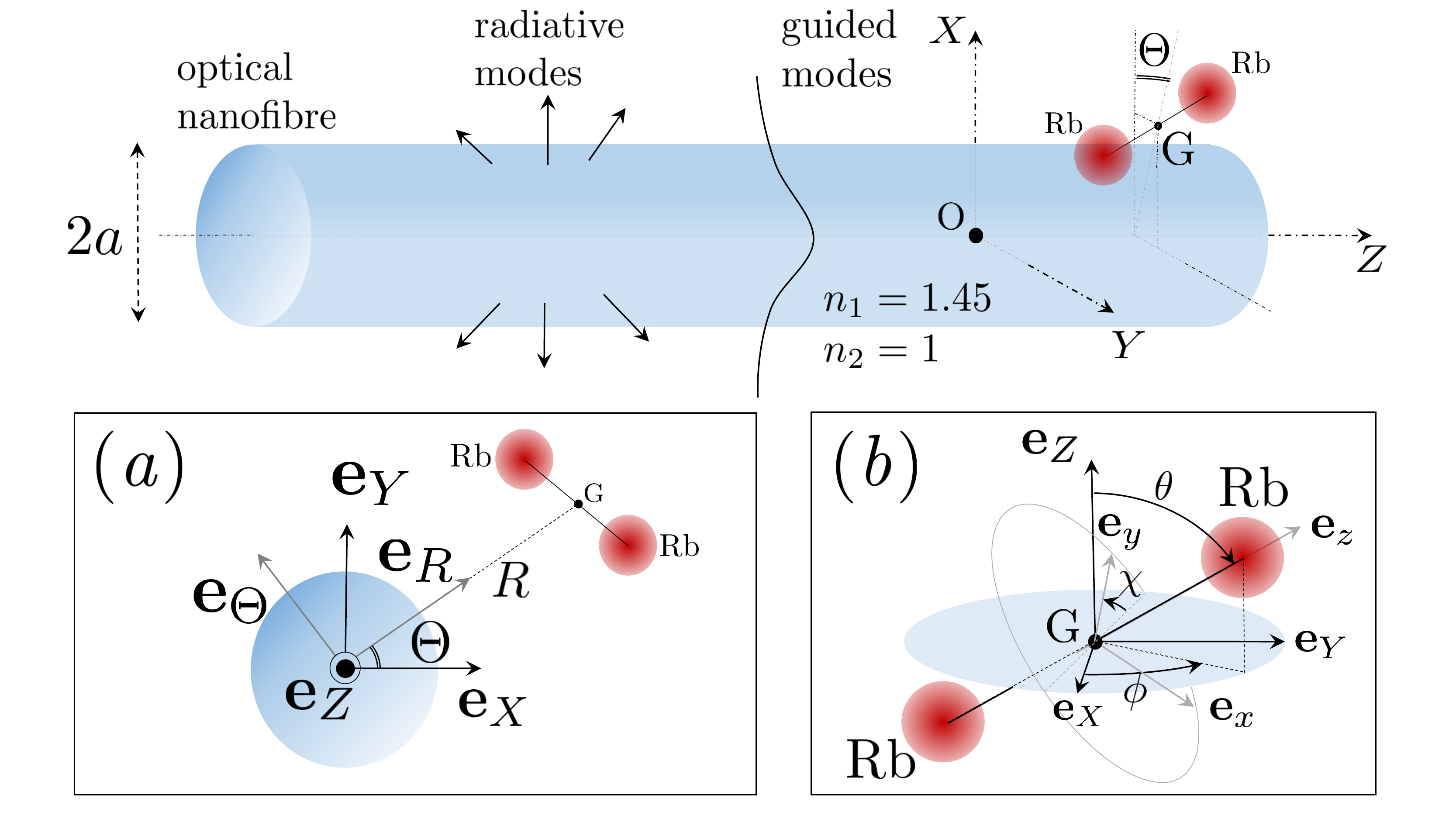}
\end{centering}
\caption{A diatomic Rb$_{2}$ molecule is located in the vicinity of an optical nanofibre
of radius $a = 200$ nm. The distance from the molecular centre of mass
to the fibre axis is denoted by $R$. Inside the fibre, the refractive
index is $n_{1} = 1.45$ (silica in the transparency window), outside
it is $n_{2}=1$ (vacuum). The axis of the nanofibre is arbitrarily
chosen as the $Z$-axis. The space-fixed Cartesian and cylindrical basis unit vectors $\left({\bf e}_{X},{\bf e}_{Y},{\bf e}_{Z}\right)$ and $\left({\bf e}_{R},{\bf e}_{\Theta},{\bf e}_{Z}\right)$ are introduced in inset (a). The position vector of the molecular centre of mass, $G$, is $R {\bf e}_R\left(\Theta \right)$. 
The molecule-fixed frame Cartesian basis unit vectors $\left({\bf e}_x, {\bf e}_y,{\bf e}_z \right)$ and Euler angles $(\theta,\phi,\chi)$ are represented in inset (b), as defined in \cite{Zare88}, pp.78-80.
}\label{FigSystem}
\end{figure*}

We consider the situation represented in figure \ref{FigSystem}. A diatomic molecule Rb$_2$ is located in the vicinity of a silica optical nanofibre of radius $a = 200$ nm, axis $(OZ)$ and whose optical index is $n_1 = 1.45$ for the frequencies of the \textsc{laser} beams we shall consider below. The centre of mass of the molecule, $G$, is at the distance $R$ from the fibre axis and the space- and molecular-fixed Cartesian frames are denoted by $(OXYZ)$ and $(Gxyz)$, respectively.

A \textsc{laser} beam of frequency $\omega_1$ is sent through the nanofibre and excites the fundamental guided mode HE$^{(X)}_{11}$ travelling along $(OZ)$ and quasi-linearly polarized along $(OX)$, of electric field positive-frequency component ${\bf E}^{(+)}_1$. Another pair of counter-propagating \textsc{laser} beams induces a standing-wave in the fundamental guided mode HE$^{(X)}_{11}$ at frequency $\omega_2 < \omega_1$, whose electric field positive-frequency component is denoted by ${\bf E}^{(+)}_{2}$. Explicit expressions for ${\bf E}^{(+)}_{1,2}$ can be found in \ref{AppGuidedMode}. These fields have a substantial evanescent component outside the fibre which interacts with the molecule. The resulting optical potential writes 
\begin{equation}
U=-\sum_{j=1,2}\left[{\bf E}_{j}^{\left(+\right)}\right]^{*}\cdot\overline{\overline{\boldsymbol{\alpha}}}\left(\omega_{j}\right)\cdot{\bf E}_{j}^{\left(+\right)}
\end{equation}
where $\overline{\overline{\boldsymbol{\alpha}}}\left(\omega \right)$ denotes the molecular dynamic polarisability tensor at frequency $\omega$ defined relatively to the state of the molecule.

The analytical expressions of the spherical components of the molecular polarisability tensor, $\alpha_{\mu \nu}$ $\left(\mu,\nu=-1,0,1\right)$,
are provided in \ref{AppenPolarizability} as sums over transitions between the states under consideration, here the rovibrational states -- denoted by $\left|\phi_n \right\rangle$ -- associated to the  $(1)^3\Sigma^+_u$ electronic state and the rovibrational states, $\left|\phi_{n'} \right\rangle$, associated to other excited electronic states, see equation (\ref{polar}). Here $n$ and $n'$ stand for complete sets of quantum numbers including, among others, $\Lambda$, the projection of the electronic orbital angular momentum onto the body-fixed $z$ axis (we recall that $\Lambda=0$ for a $\Sigma$ electronic state, $\Lambda=\pm1$ for $\Pi$ states), as well as $v$ and $J$ the vibrational and rotational quantum numbers, respectively. In practice, we restricted ourselves to 
the two perpendicular ($\Lambda-\Lambda'=\pm1$) transitions $(1)^3\Sigma^+_u \rightarrow (1)^3\Pi_g$ and $(1)^3\Sigma^+_u \rightarrow (2)^3\Pi_g$, and the two parallel  ($\Lambda-\Lambda'=0$) transitions  $(1)^3\Sigma^+_u \rightarrow (1)^3\Sigma_g$ and  $(1)^3\Sigma^+_u \rightarrow (2)^3\Sigma_g$. We included 500 vibrational states in the calculations. The corresponding electronic transition dipole moments are given in \cite{AA12}. 
In  the present case, because of the geometry of the molecule, only three spherical components of the polarisability tensor are non zero : $\alpha_{-1 1}$, $\alpha_{1 -1}$ and $\alpha_{0 0}$. From them, the three diagonal Cartesian components of the polarisability tensor are obtained: $\alpha_{X X}=\alpha_{Y Y}=-\frac12 \left( \alpha_{1 -1}+\alpha_{-1 1}\right)$, $\alpha_{Z Z}=\alpha_{00}$. These Cartesian components provide a simple geometrical interpretation of the results.   

\begin{figure*}
\begin{centering}
\includegraphics[
width=\textwidth]{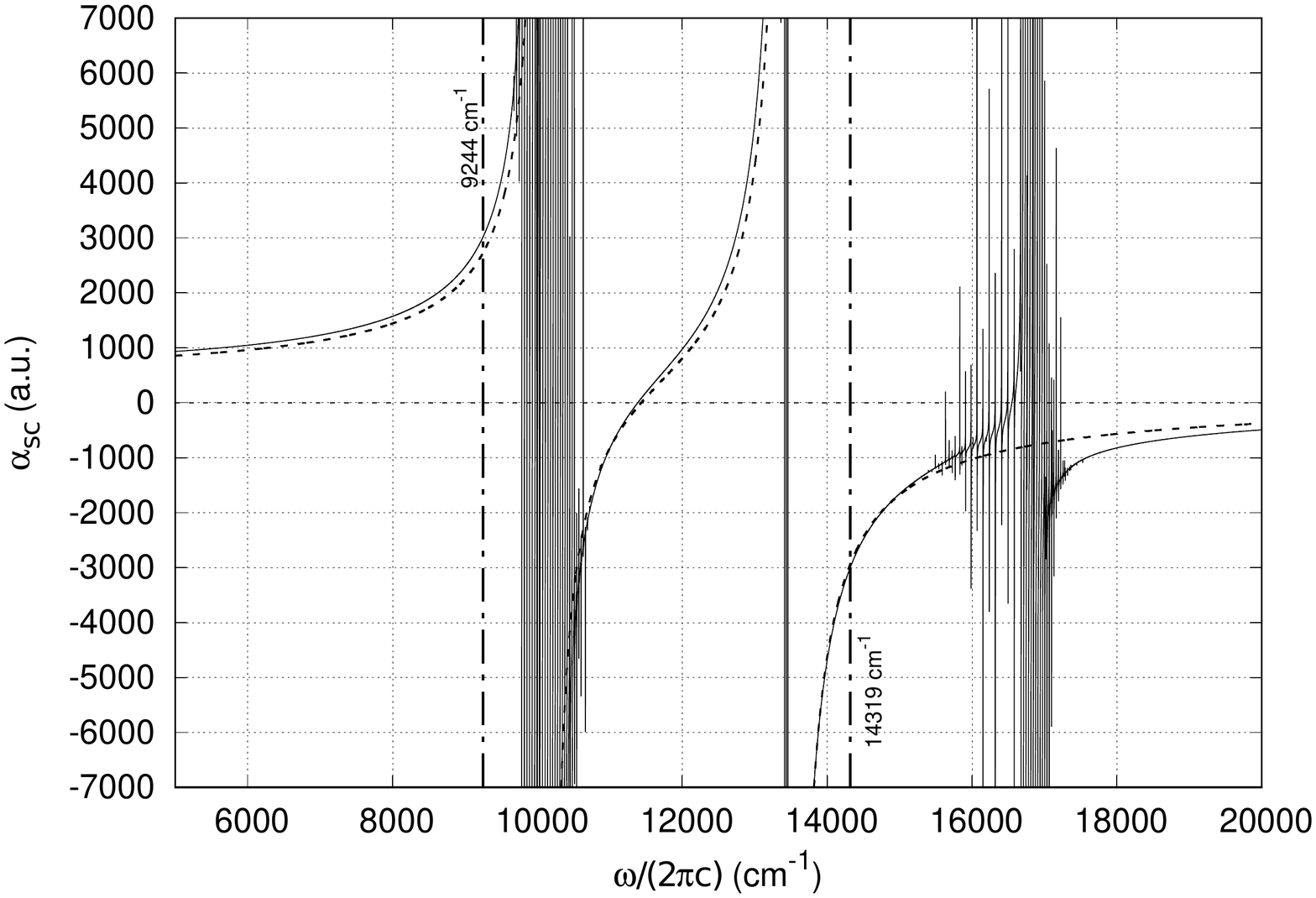}
\end{centering}
\caption{
Full line : Scalar polarisability of Rb$_2$($(1)^3\Sigma^+_u$) as a function of field wavenumber, from our numerical simulation. Dashed line : the scalar polarisability from the analytical fit given by equation (8) in \cite{DDD15}. The two vertical dashed-dotted lines correspond to the wavenumbers of the two fields used to trap the molecule. 
} \label{PolarFig}
\end{figure*}

It is known that the best representation of the Rb$_2$ molecule is provided by Hund's case (b) \cite{PKH17,STL10}. Then additional quantum numbers -- the total angular momentum $J$, $M$ its projection on the $Z$ axis as well as $N$, the total angular momentum exclusive of spins -- are necessary  to fully characterize the molecular state. In this case,  the polarisability of the ground rovibrational state $\left| (1)^3\Sigma^+_u, v=0, J=1, N=0,M=0,\pm1\right\rangle$ is independent of $M$ and purely scalar (neither vector nor tensor components). The scalar part given by
\[
\label{polarsca}
\alpha_{\mbox{sc}}=\frac13
\sum_{\mu=-1,0,1} (-1)^\mu \alpha_{\mu -\mu}
\]
is shown on fig.  \ref{PolarFig} as a function of  $\omega$. 
The polarisability goes through divergences in resonance bands associated to the transitions between rovibrational states supported by the $^3\Sigma_g$ and $^3\Pi_g$ potentials.
For the sake of comparison, we also plotted the approximate analytical form given by equation (8) in \cite{DDD15}. The latter was obtained from a fit of another numerical calculation performed with different quantum chemistry data. In particular, the numerical simulation in \cite{DDD15} took more excited electronic states into account. Except in the resonance bands -- where the fit is anyway not expected to be valid -- we observe a good agreement between our results and those of \cite{DDD15}. This suggests that the restricted set of electronic states considered in our calculations is sufficient to get accurate values for the polarisability.

We note that the polarisability is positive for frequencies lower than those in the resonance band and negative above, as expected from the presence of factors $\nicefrac{1}{\left( \omega_{n'n}^2-\omega^2 \right)}$ in equation (\ref{polar}). Below we shall take advantage of this property to design a bichromatic optical trapping potential.

Although it is not adapted to the molecule at stake, we shall also consider Hund's case (a). In this limit, polarisability has specific alignment properties and recovers its full tensorial character. The projection of the spin angular momentum onto the molecular-fixed $z$ axis, $\Sigma=0, \pm1$, now becomes a good quantum number and so does the total angular momentum projection, $\Omega=\Lambda+\Sigma$. The wavefunction has specific rotational properties which reflect in the polarization alignment properties. Treating the particular Hund's case (a) goes beyond a purely textbook-like exercise since it will allow us to draw general conclusions on how the trapping of a molecule by an anisotropic electric field is influenced by its alignment properties. 

\section{Numerical results and discussion\label{Results}}

In this section, we present the numerical results we obtained. First, we specify the physical parametres we considered for the \textsc{laser} fields and the molecular polarisability tensor components at the relevant frequencies. We then analyse the trapping potentials we calculated for the molecule treated in either  Hund's cases (a) or (b) and specify their main features. Finally, we discuss some possible limitations and perspectives of our calculations.    

\subsection*{System parametres}
Table \ref{Tabparametres} provides the parametres of the travelling and standing-wave fields: the (absolute) frequency, $\omega$, the adimensioned wavevector in vacuum, $k a$, the adimensioned propagation constant, $\beta a$, solution of the characteristic equation (\ref{eveq}), the (adimensioned) parametres $ha$ and $qa$ as defined in equation (\ref{eqhq}), the parametre $s$ as defined in equation (\ref{eqs}), the amplitude, $\mathcal{A}$, and power, $\Pi$, in atomic units. Note that the standing-wave field intensity is periodic in $Z$ and the period is given by $\pi / \beta_2 \approx 530$ nm.

\begin{table}
\begin{centering}
\begin{tabular}{|c|c|c|}
\hline 
 & \textbf{Travelling field} & \textbf{Standing-wave field}\tabularnewline
\hline 
\hline 
$\omega$  & $\omega_{1}=14319\mbox{ cm}^{-1}$ & $\omega_{2}=9244\mbox{ cm}^{-1}$\tabularnewline
\hline 
$ka$ & $k_{1}a\approx1.79938$ & $k_{2}a\approx1.16164$\tabularnewline
\hline 
$\beta a$ & $\beta_{1}a\approx2.03575$ & $\beta_{2}a\approx1.18227$\tabularnewline
\hline 
$ha$ & $h_{1}a\approx1.62297$ & $h_{2}a\approx1.19467$\tabularnewline
\hline 
$qa$ & $q_{1}a\approx0.952109$ & $q_{2}a\approx0.219907$\tabularnewline
\hline 
$s$ & $s_{1}\approx-0.856465$ & $s_{2}\approx-0.971884$\tabularnewline
\hline 
$\mathcal{A}$ & $\mathcal{A}_{1}\approx2.54951\times10^{-6}$ a.u. & $\mathcal{A}_{2}\approx 8\times10^{-7}$ a.u.\tabularnewline
\hline 
$\Pi$ & $\Pi_{1}\approx6.80693\times10^{-6}$ a.u. & $\Pi_{2}\approx4.30252\times10^{-6}$ a.u.\tabularnewline
\hline 
\end{tabular}
\end{centering}
\caption{Physical parametres of the trapping \textsc{laser} fields (see main text for
the definition of the quantities listed here). 
}
\label{Tabparametres}
\end{table}

On the other hand, Table \ref{TabPolar} shows the non-vanishing components of the molecular polarisability tensor for the Rb$_2$ molecule prepared in the metastable state $(1)^3 \Sigma_u^{+}$ at the \textsc{laser} frequencies $\omega_{1,2}$. The values were numerically calculated in the two limiting Hund's cases (a) and (b). In the latter case, it is restricted to its scalar component and it does not depend on the value of $M=0,\pm 1$.
These results can be interpreted in terms of the alignment properties of the molecular states, as we shall now see. The molecular polarisability results from the combined contributions of $\Sigma \rightarrow \Sigma$ and $\Sigma \rightarrow \Pi$ transitions, whose moments are respectively parallel and orthogonal to the molecular axis. To be more explicit, the space-fixed Cartesian components of the polarisability are given by \cite{DDD14}
\begin{equation}
\label{align}
\alpha_{ii} = a_i  \alpha_{\parallel} + \left(1- a_i \right) \alpha_{\perp}
\end{equation}
where $i=X, Y, Z$, $\alpha_{\parallel,\perp}$ denote the respective contributions of $\Sigma \rightarrow \Sigma$ and $\Sigma \rightarrow \Pi$ transitions, and the $a_i$'s quantify the degree of alignment of the molecular axis ${\bf e}_z$ with respect to the space axes for a given space-fixed molecular state $|\phi_n^s \rangle$ (see \ref{AppenPolarizability})
\begin{equation}
\label{ai}
a_i=\left\langle\phi_n^s\right|\left({\bf e}_i.{\bf e}_z\right)^2 \left|\phi_n^s\right\rangle
\end{equation}
with the obvious relation $\sum_i a_i=1$. 
If the molecular were laid isotropically in space, one would have $a_i=\frac13$. Deviation from this value is therefore indicative of anisotropy in the molecular layout. The alignment parametres extracted from the polarizabilities in Table \ref{TabPolar} using equation \ref{align} are given in the same table.
As expected, only the Hund's case (b) state, $N=0$, is fully isotropic. Note that the alignment parametres $a_i$'s can also be obtained directly from equation \ref{ai} by integration over the angles $\left(\theta, \phi\right)$ (figure \ref{FigSystem}), since the states $\left|\phi_n^s\right\rangle$ depend on these angles through the symmetric top rotational wavefunctions $\left|J \Omega M \right\rangle$ (equation \ref{phimol}), which are known analytically as Wigner rotation matrix elements (see \cite{Zare88}, equation (3.125), p. 105). 
The alignement parametres 
extracted from the polarisability tensor coincide with 
those obtained by direct analytical angular integration.

\begin{table}
\begin{centering}
\begin{tabular}{|c|c|c|}
\cline{2-3} \cline{3-3} 
\multicolumn{1}{c|}{} & $\omega_{1}$ & $\omega_{2}$\tabularnewline
\cline{2-3} \cline{3-3} 
\multicolumn{1}{c|}{} & $\begin{array}{c}
\alpha_{\mbox{sc}}=-3000.92\\
\alpha_{\parallel}=-1034.53\\
\alpha_{\perp}=-3984.11
\end{array}$ & $\begin{array}{c}
\alpha_{\mbox{sc}}=2998.83\\
\alpha_{\parallel}=6804.32\\
\alpha_{\perp}=1096.07
\end{array}$\tabularnewline
\hline 
\hline 
\multicolumn{3}{|c|}{\textbf{Hund's case (a)}}\tabularnewline
\hline 
\multirow{2}{*}{$\left(M,\Sigma\right)=\left\{ \begin{array}{c}
\left(0,\pm1\right)\\
\left(\pm1,0\right)
\end{array}\right.$} & $\begin{array}{c}
\alpha_{XX}=\alpha_{YY}=-2804.28\\
\alpha_{ZZ}=-3394.20
\end{array}$ & $\begin{array}{c}
\alpha_{XX}=\alpha_{YY}=3379.37\\
\alpha_{ZZ}=2237.73
\end{array}$\tabularnewline
\cline{2-3} \cline{3-3} 
 & \multicolumn{2}{c|}{$a_{X}=a_{Y}=0.4\qquad a_{Z}=0.2$}\tabularnewline
\hline 
\multirow{2}{*}{$\left(M,\Sigma\right)=\left(0,0\right)$} & $\begin{array}{c}
\alpha_{XX}=\alpha_{YY}=-3394.20\\
\alpha_{ZZ}=-2214.36
\end{array}$ & $\begin{array}{c}
\alpha_{XX}=\alpha_{YY}=2237.73\\
\alpha_{ZZ}=4521.02
\end{array}$\tabularnewline
\cline{2-3} \cline{3-3} 
 & \multicolumn{2}{c|}{$a_{X}=a_{Y}=0.2\qquad a_{Z}=0.6$}\tabularnewline
\hline 
\multirow{2}{*}{$\left(M,\Sigma\right)=\left\{ \begin{array}{c}
\left(1,\pm1\right)\\
\left(-1,\pm1\right)
\end{array}\right.$} & $\begin{array}{c}
\alpha_{XX}=\alpha_{YY}=-3099.24\\
\alpha_{ZZ}=-2804.28
\end{array}$ & $\begin{array}{c}
\alpha_{XX}=\alpha_{YY}=2808.55\\
\alpha_{ZZ}=3379.37
\end{array}$\tabularnewline
\cline{2-3} \cline{3-3} 
 & \multicolumn{2}{c|}{$a_{X}=a_{Y}=0.3\qquad a_{Z}=0.4$}\tabularnewline
\hline 
\hline 
\multicolumn{3}{|c|}{\textbf{Hund's case (b)}}\tabularnewline
\hline 
\multirow{2}{*}{$\begin{array}{c}
M=0,\pm1\\
N=0
\end{array}$} & $\alpha_{XX}=\alpha_{YY}=\alpha_{ZZ}=-3000.92$ & $\alpha_{XX}=\alpha_{YY}=\alpha_{ZZ}=2998.83$\tabularnewline
\cline{2-3} \cline{3-3} 
 & \multicolumn{2}{c|}{$a_{X}=a_{Y}=a_{Z}=\nicefrac{1}{3}$}\tabularnewline
\hline 
\end{tabular}
\par\end{centering}
\caption{Scalar, parallel, perpendicular polarizabilities and non vanishing space-fixed Cartesian components of the molecular polarisability tensor
$\left\{ \alpha_{XX},\alpha_{YY},\alpha_{ZZ}\right\} $  
in the Hund's cases $\left(a,b\right)$ for different $\left(M,\Sigma\right)$ and $N$ quantum numbers. The alignment parametres $a_X,a_Y,a_Z$ are also given. The molecule is in the state $(1)^3\Sigma^+_u, v=0, J=1$. The polarizabilities are in atomic units.}
\label{TabPolar}
\end{table}

\subsection*{Trapping potential}
\begin{figure}
    \centering
     \begin{subfigure}[b]{\textwidth}
         \centering
         \includegraphics[width=5cm]{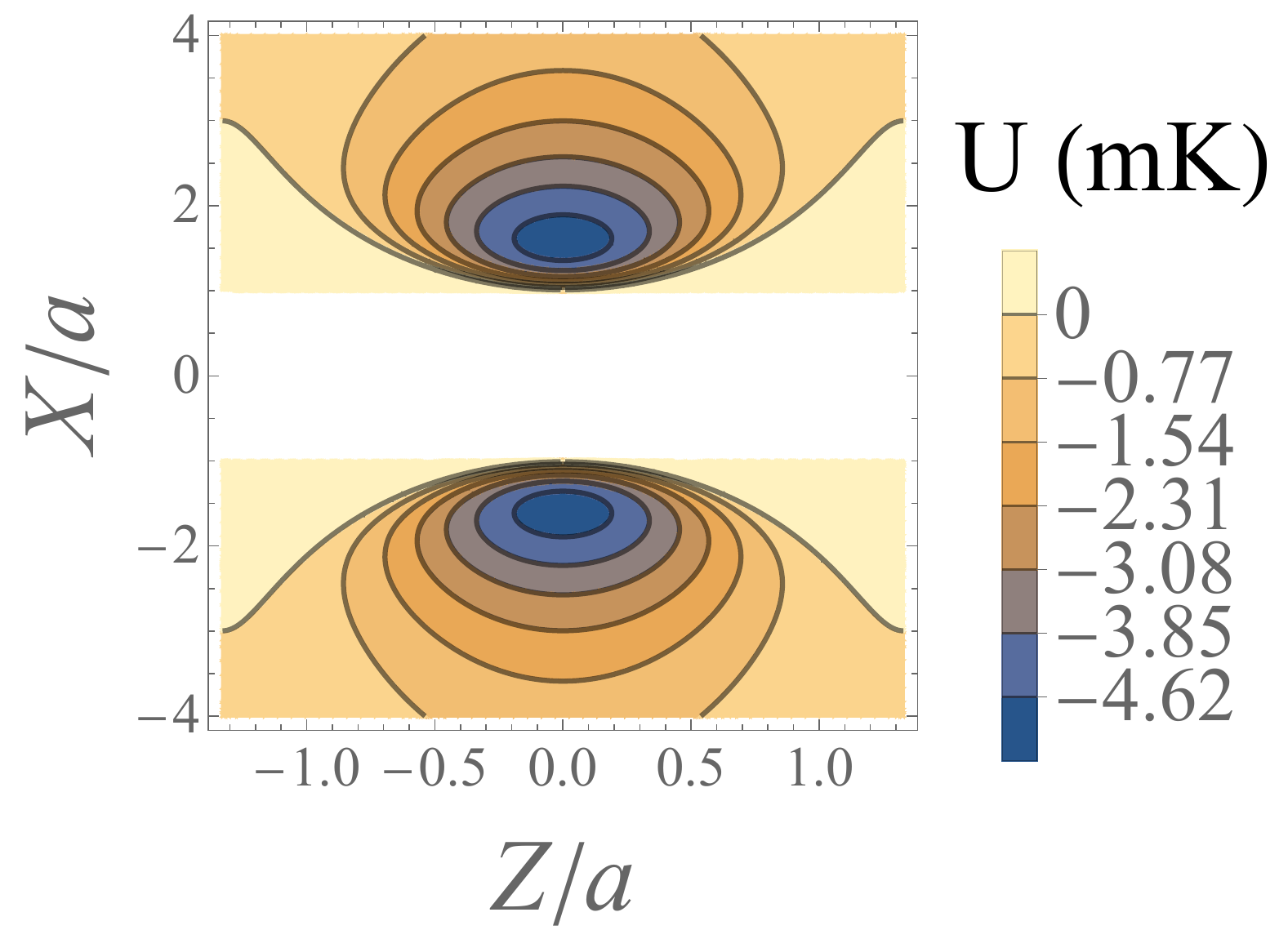}
         \hfill
         \includegraphics[width=5cm]{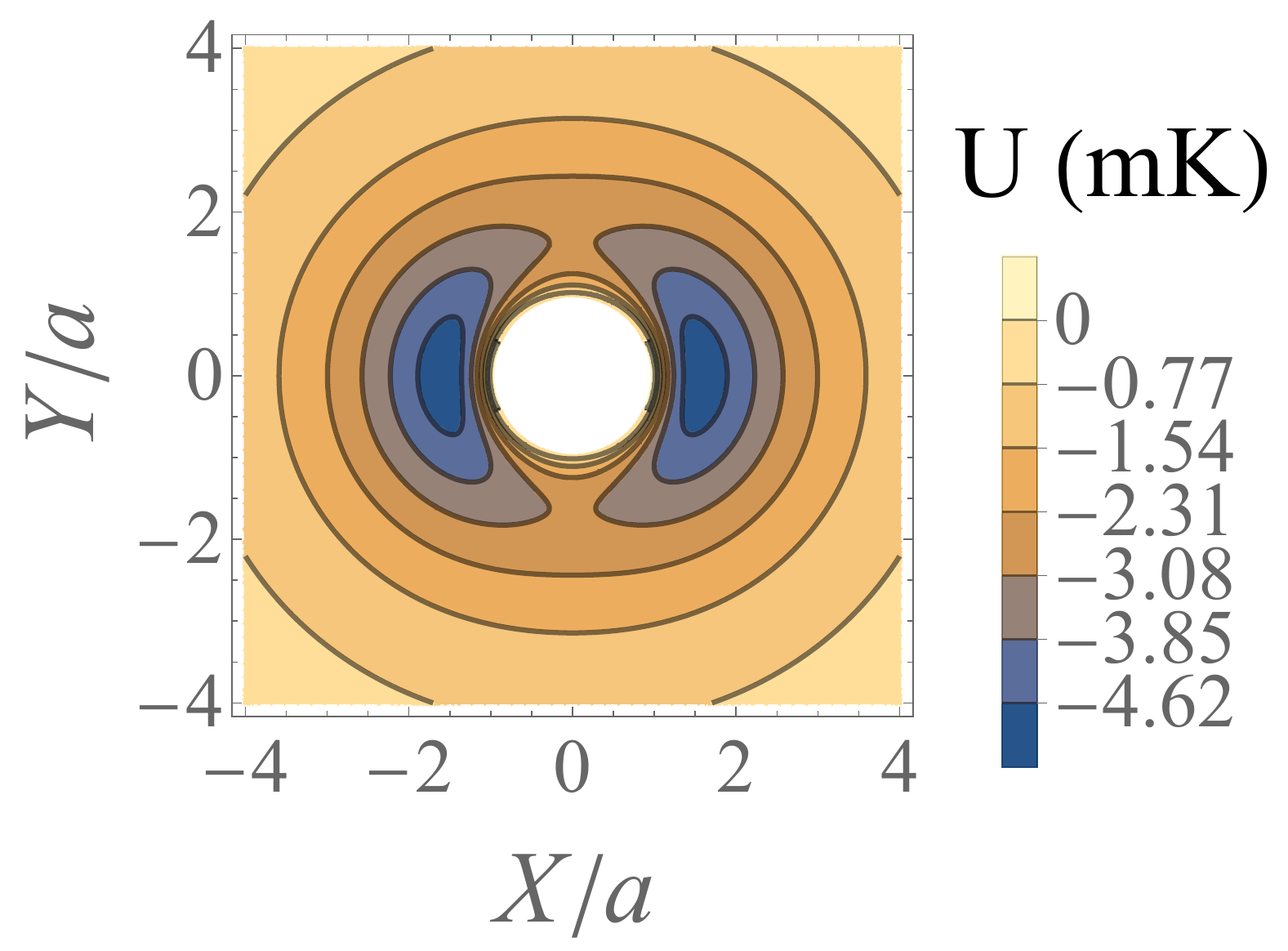}
         \caption{Hund's case (a), $(M,\Sigma)=(0,\pm 1),(\pm 1,0)$}
     \end{subfigure}
     \\
     \centering
     \begin{subfigure}[b]{\textwidth}
         \centering
         \includegraphics[width=5cm]{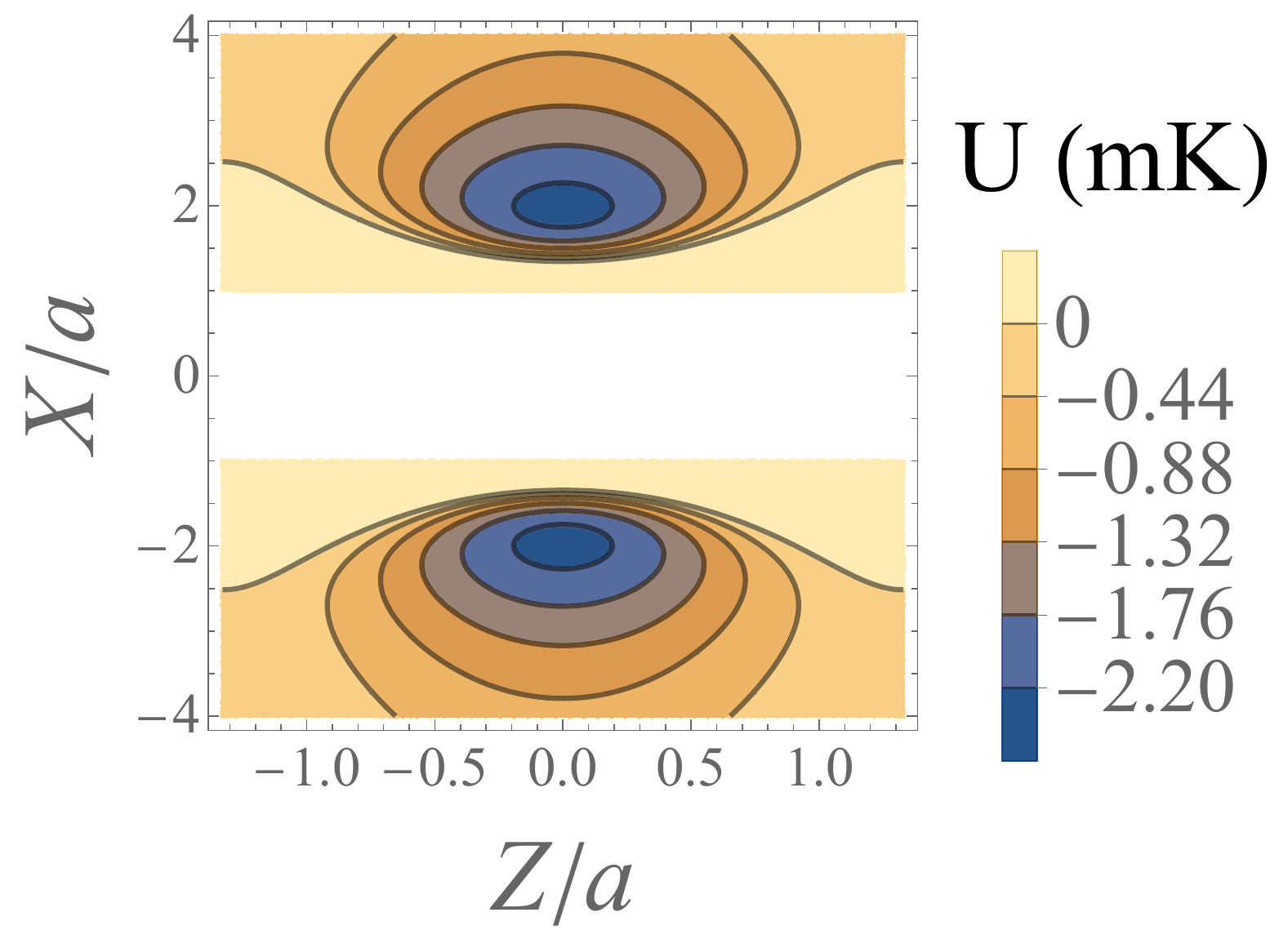}
         \hfill
         \includegraphics[width=5cm]{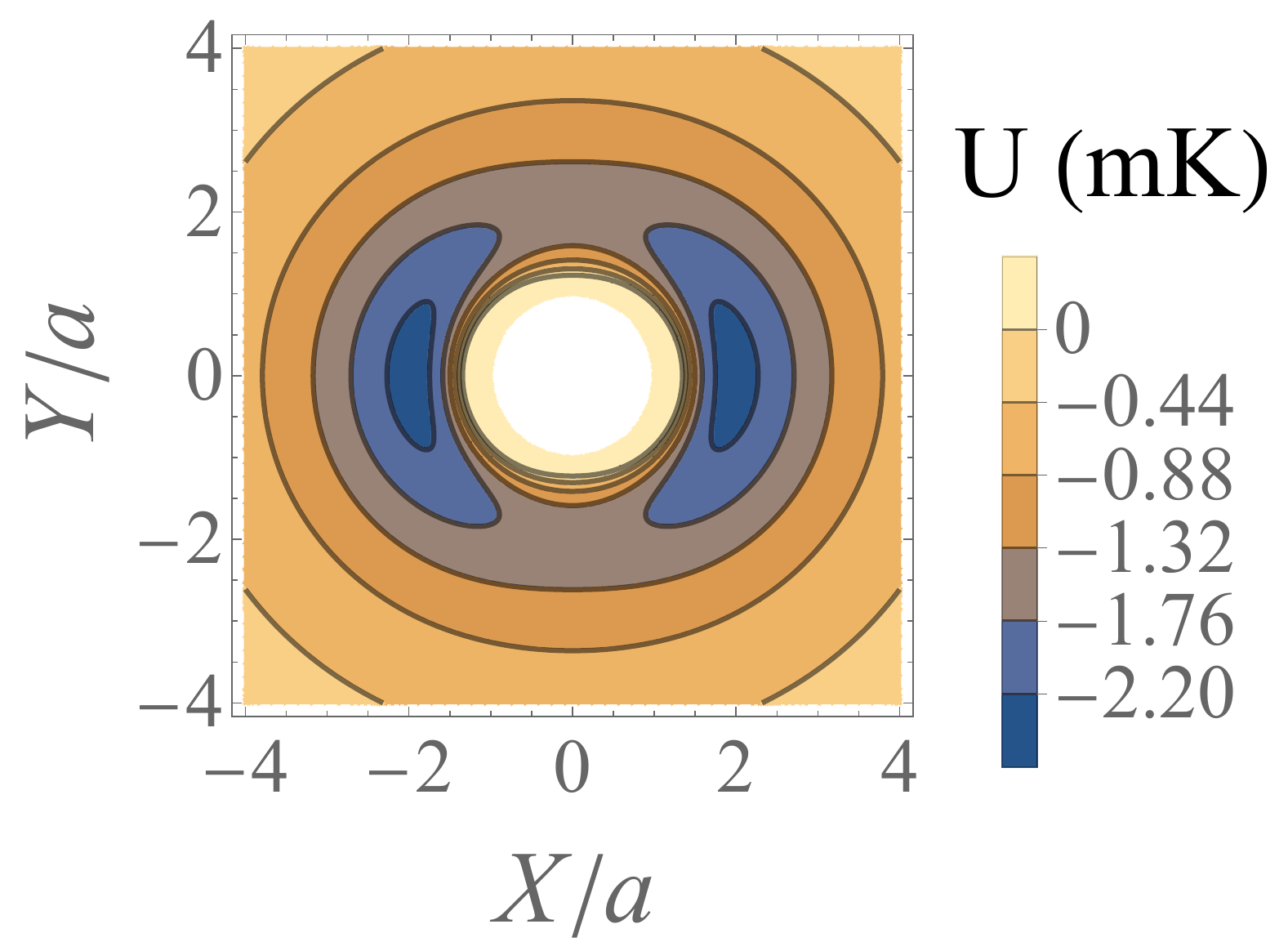}
         \caption{Hund's case (a), $M=\Sigma=0$}
     \end{subfigure}
      \\
        \centering
     \begin{subfigure}[b]{\textwidth}
         \centering
         \includegraphics[width=5cm]{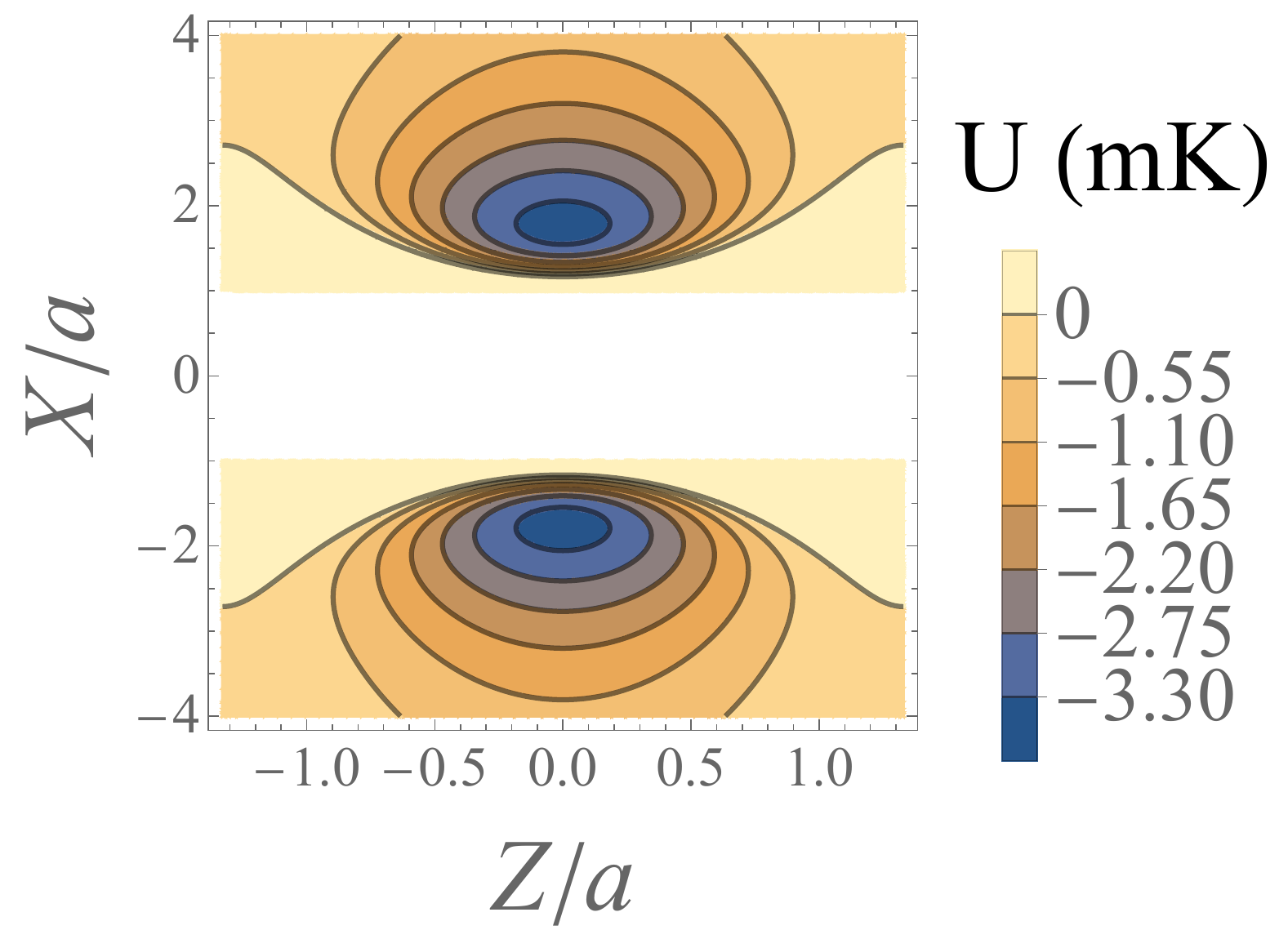}
         \hfill
         \includegraphics[width=5cm]{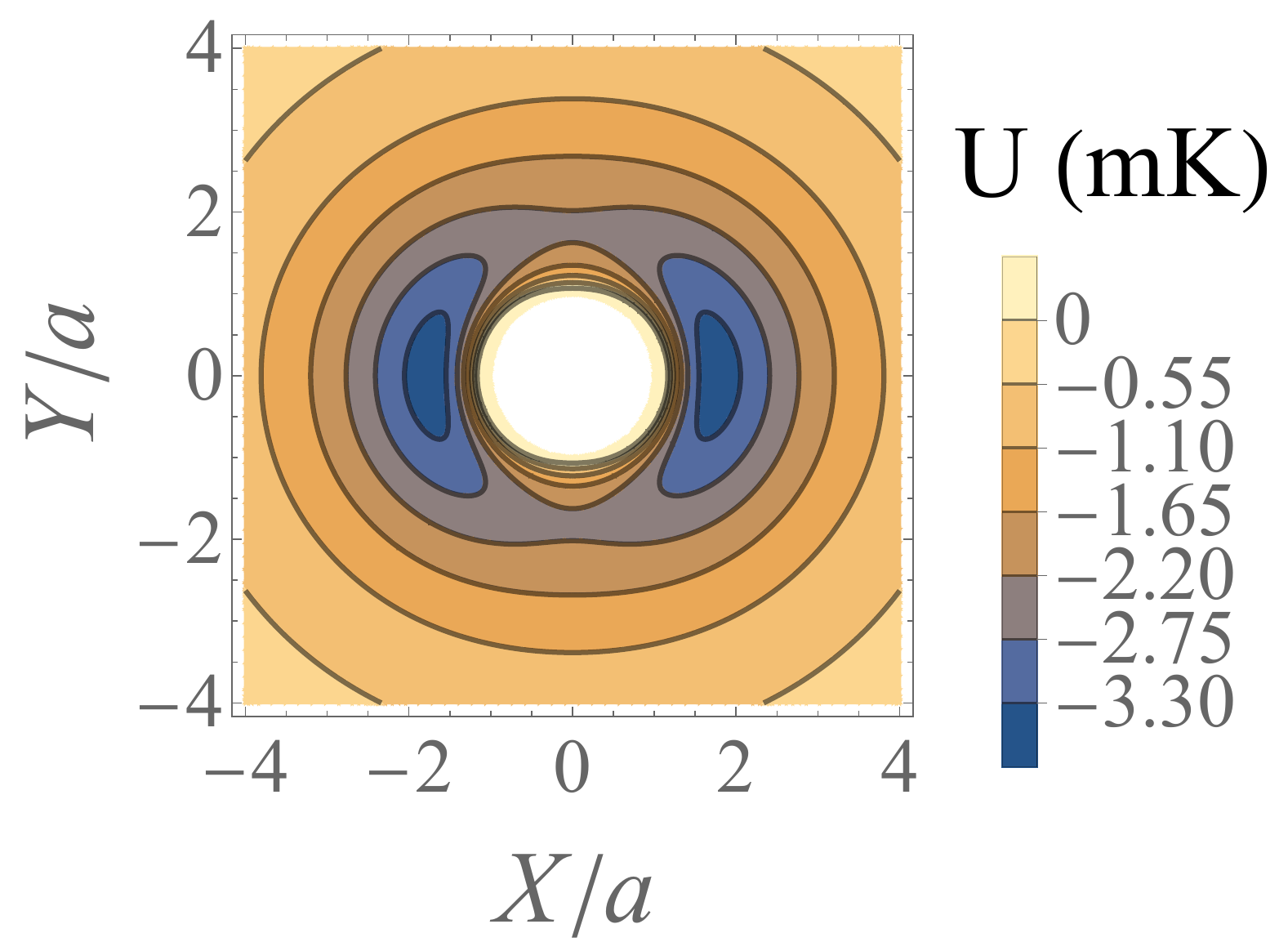}
         \caption{Hund's case (a), $(M,\Sigma)=(1,\pm 1),(- 1, \pm 1)$}
     \end{subfigure}
     \\
     \centering
     \begin{subfigure}[b]{\textwidth}
         \centering
         \includegraphics[width=5cm]{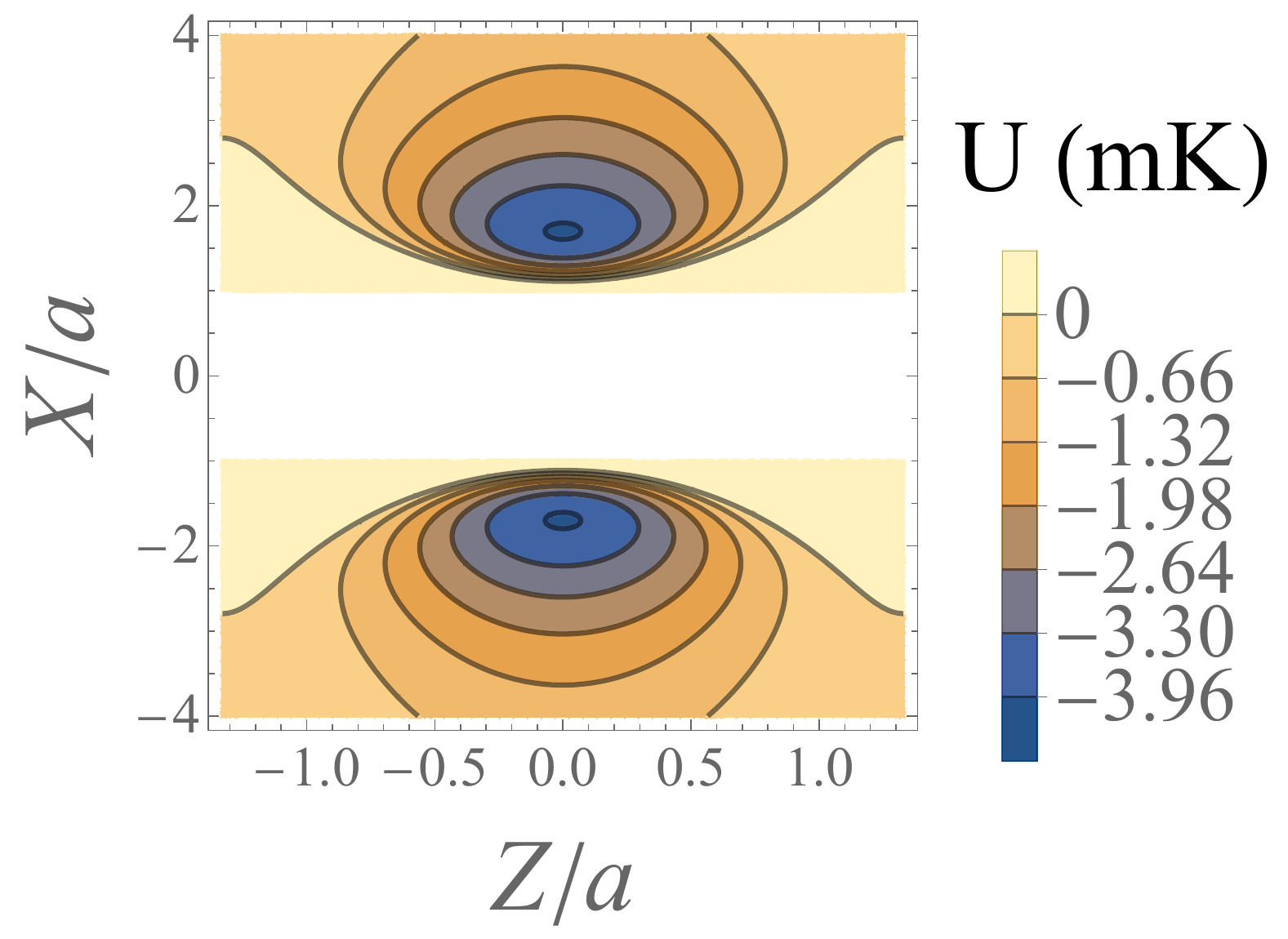}
         \hfill
         \includegraphics[width=5cm]{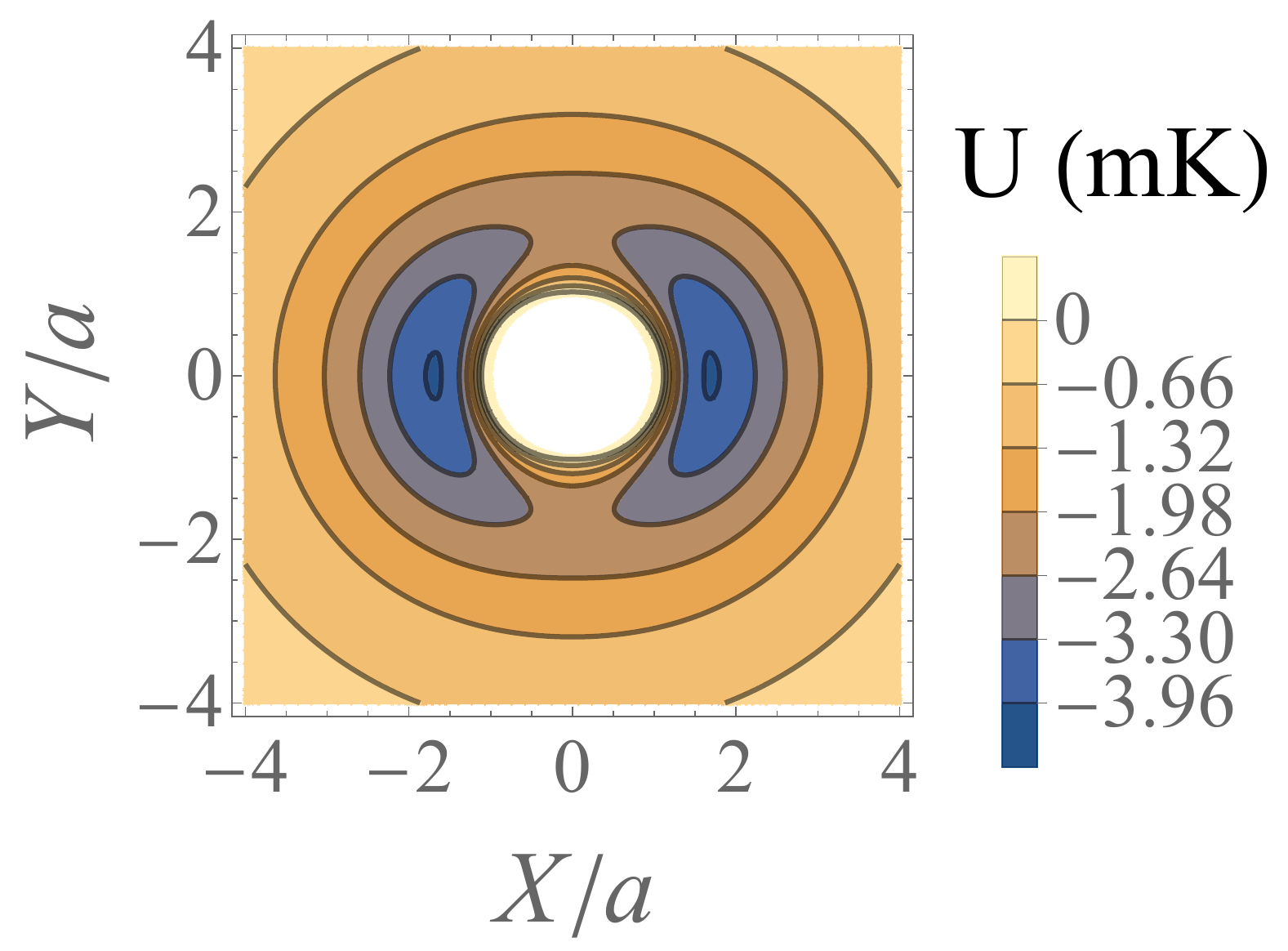}
         \caption{Hund's case (b), $M=-1,0,1$}
     \end{subfigure}
     
        \caption{Contour plots of the trapping potential $U\left( X,Y,Z \right)$ (in mK) as a function
of $\left( X,Z \right)$ for $Y=0$ (left column) and $\left( X,Y \right)$ for $Z=0$ (right column) for Hund's cases (a) (subfigures a,b,c) and (b) (subfigure d). In the left column, we restricted the $Z$ range to one period of the potential, i.e. one period of the standing-wave guided field $ \left(\pi / \beta_2 \right )\approx 530$ nm.}
\label{contourplots}       
\end{figure}

\begin{figure}
     \centering
     \begin{subfigure}[b]{\textwidth}
         \centering
         
         \caption{}
         
         \includegraphics[width=\textwidth]{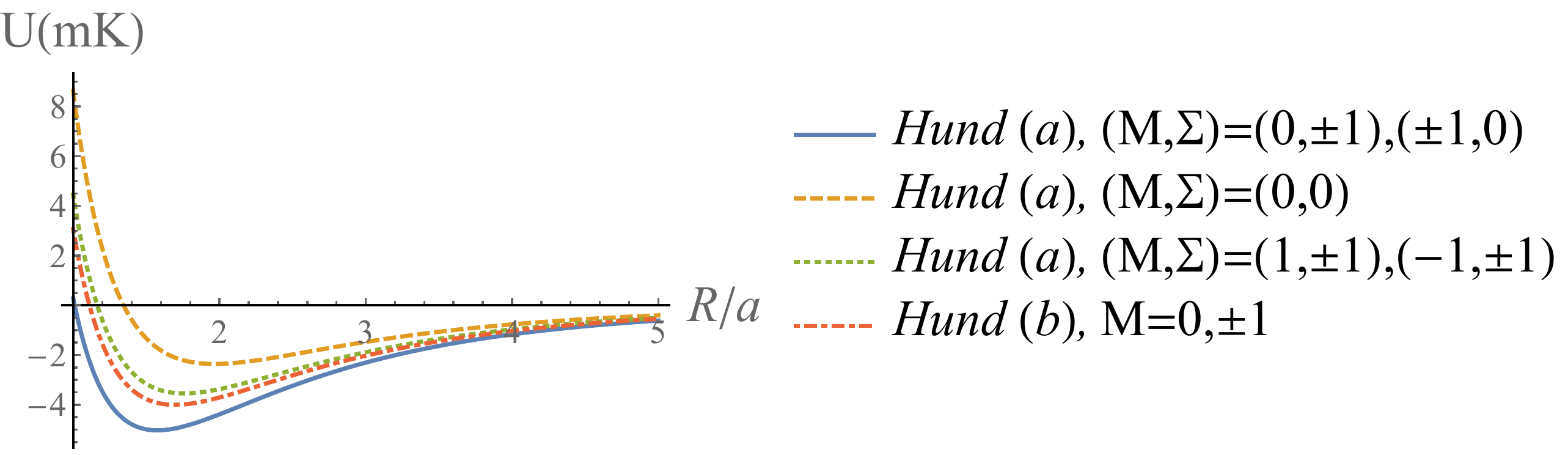}
     \end{subfigure}
     \vspace{1cm}
     \\
     \begin{subfigure}[b]{\textwidth}
        \centering
        \caption{}
        \includegraphics[width=\textwidth]{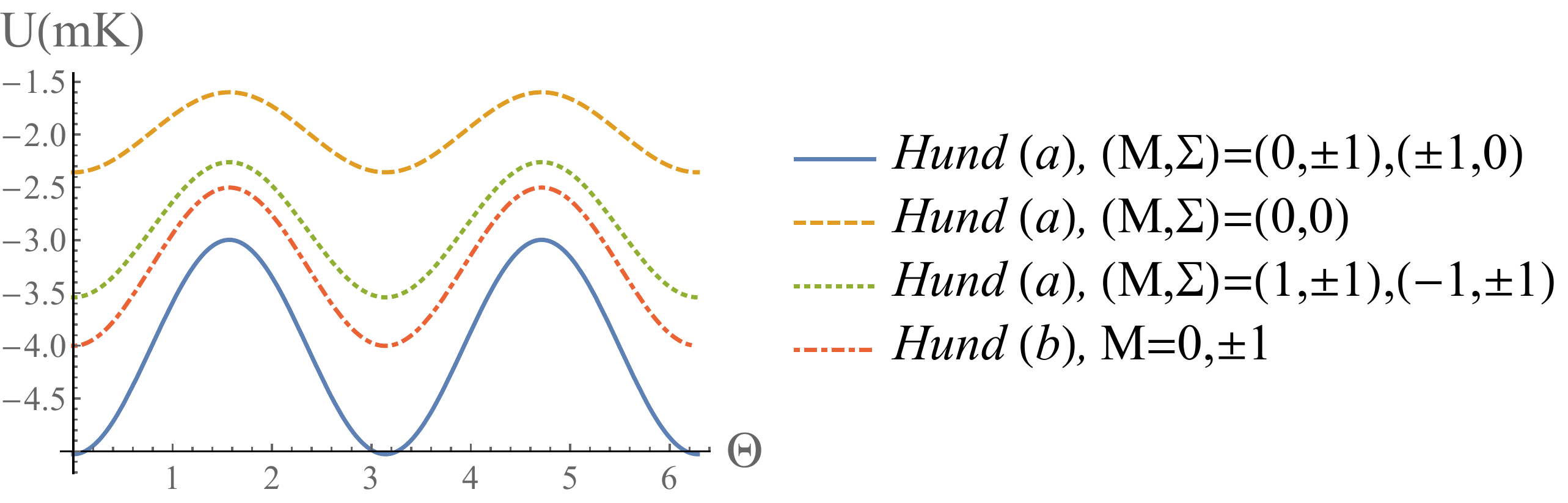}
     \end{subfigure}
     \vspace{1cm}
     \\
     \begin{subfigure}[b]{\textwidth}
         \centering
         \caption{}
         \includegraphics[width=\textwidth]{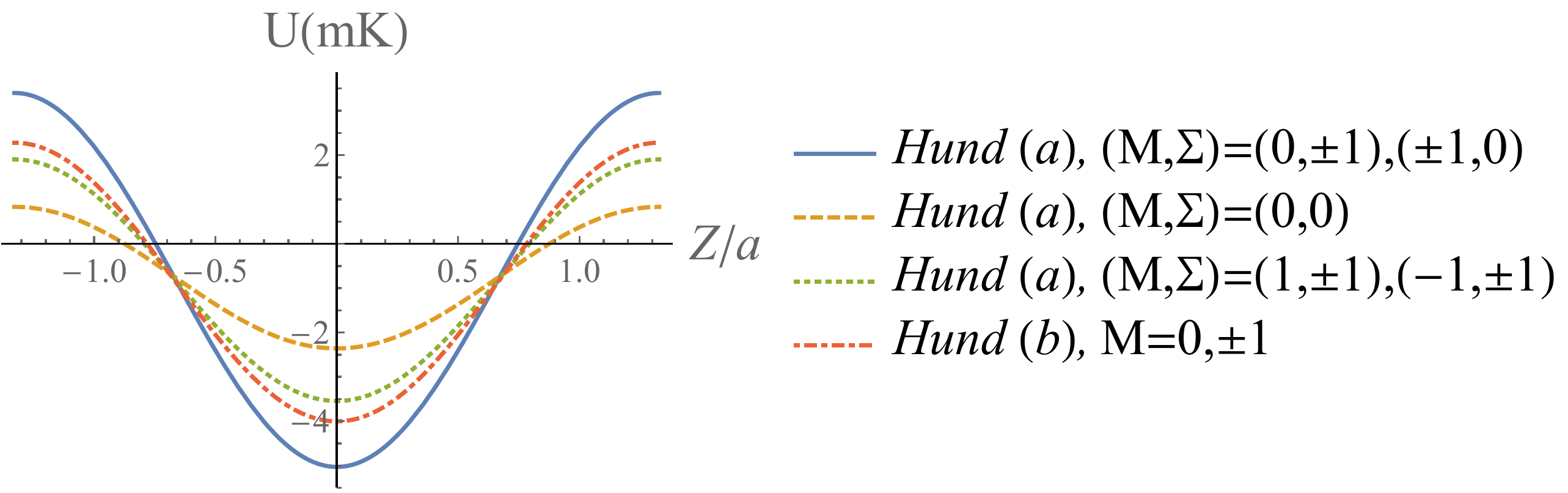}
     \end{subfigure}
        \caption{Trapping potential $U\left(R,\Theta,Z\right)$ (in mK) for Hund's cases (a,b) as a function
of (a) $R$ for $Z=0$ and $\Theta=0$, (b) $\Theta$ for $R=R_{\mbox{min}}^{(s)}$ and $Z=0$, (c) $Z$ for $R=R_{\mbox{min}}^{(s)}$ and $\Theta=0$. Here $R_{\mbox{min}}^{\left(s\right)}$ denotes the value of $R$ for which the trapping potential is minimal when $\Theta=0$ and $Z=0$ for the state $s$ (defined by the Hund's case and $M, \Sigma$), as determined on figure (a).
}\label{Potentialrhophiz}
\end{figure}

Figure \ref{contourplots} shows contour plots of the two-lobe trapping potential $U$ we obtained in Hund's cases (a) (subfigures a, b, and c), and (b) (subfigure d) as a function of $(X,Z)$ for $Y=0$ (left column) and $(X,Y)$ for $Z=0$. This is complemented by figure \ref{Potentialrhophiz} which shows the behaviour of $U$ as a function of $R$ when $\Theta=0$ and $Z=0$ (subfigure a), $\Theta$ when $Z=0$ and $R$ is set to the value $R_{\mbox{min}}^{(s)}$ which minimizes the potential when $\Theta=0$ and $Z=0$ for the state $s$ considered  (subfigure b), and $Z$ when $R=R_{\mbox{min}}^{(s)}$ and $\Theta=0$ (subfigure c).

For the specific Hund's case (b), a potential minimum $U_{\mbox{min}} \approx - 4$~mK is obtained for $R\approx 1.7 a \approx 340$ nm,  $\Theta = 0,\pi$ and $Z = \nu \pi /\beta_2$ with $\nu \in \mathbb{Z}$ (subfigure  \ref{contourplots} d). 
In this case, the trapping potential is given by : 
$U = - \sum_{j=1,2} \alpha_{\mbox{sc}}^{(j)} \left|{\bf E}_j^{(+)}\right|^2$, 
where $\alpha_{\mbox{sc}}^{(j=1,2)}$ denote the scalar polarizabilities at the travelling and standing-wave field frequencies, respectively. 
Since  $\alpha_{\mbox{sc}}^{(1)} \approx - \alpha_{\mbox{sc}}^{(2)}$ (see Table \ref{TabPolar}), 
one has $U \approx  - \alpha_{\mbox{sc}}^{(2)} \left( \left|{\bf E}_2^{(+)}\right|^2 - \left|{\bf E}_1^{(+)}\right|^2 \right)$ 
and the trap is located in the region of space where 
$\left|{\bf E}_2^{(+)}\right|^2 - \left|{\bf E}_1^{(+)}\right|^2$ is maximal.

\begin{figure}
    \centering
     \begin{subfigure}[b]{\textwidth}
         \centering
         
         \caption{}
         
         \includegraphics[width=6cm]{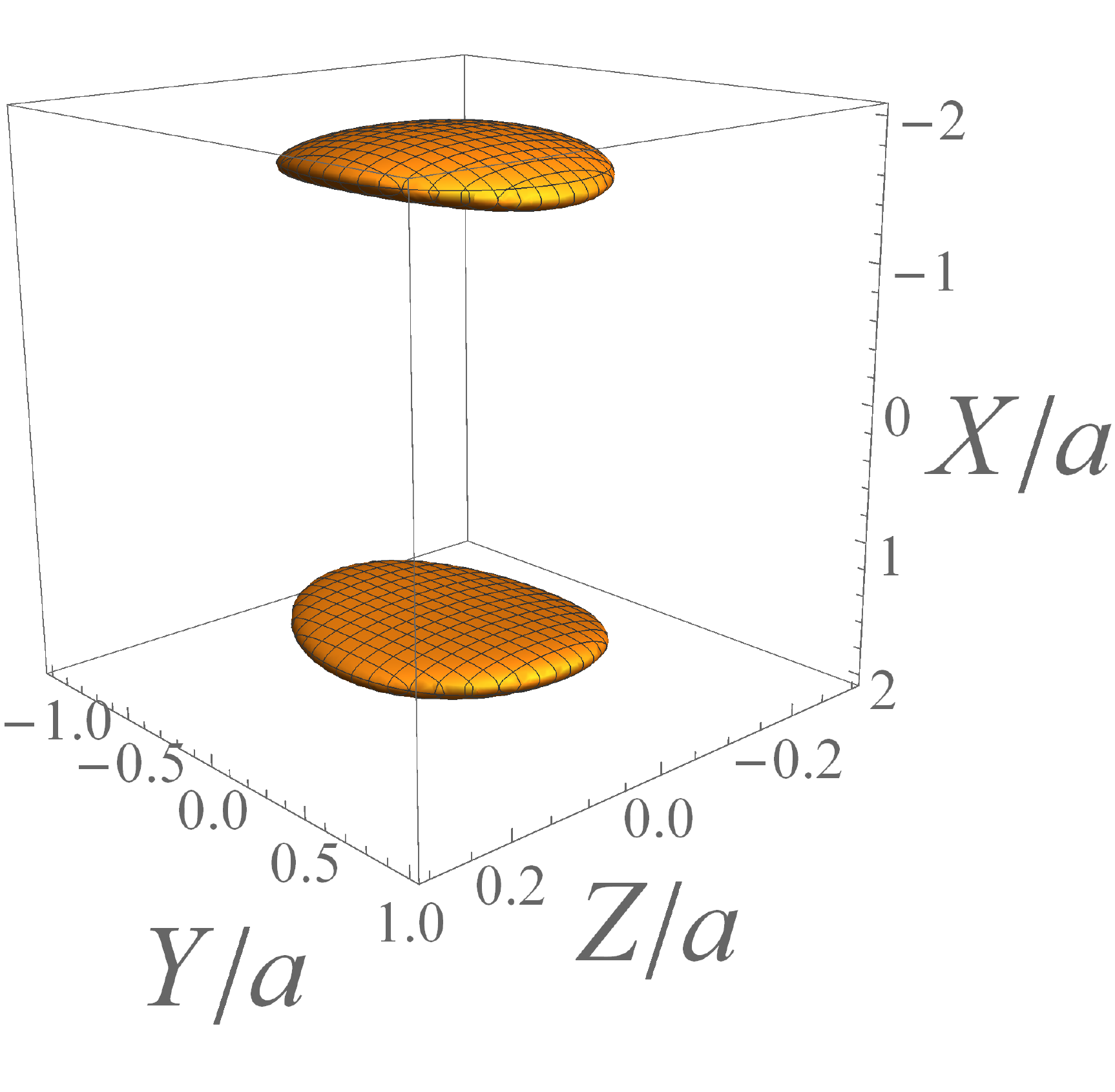}
         \hfill
         \includegraphics[width=6cm]{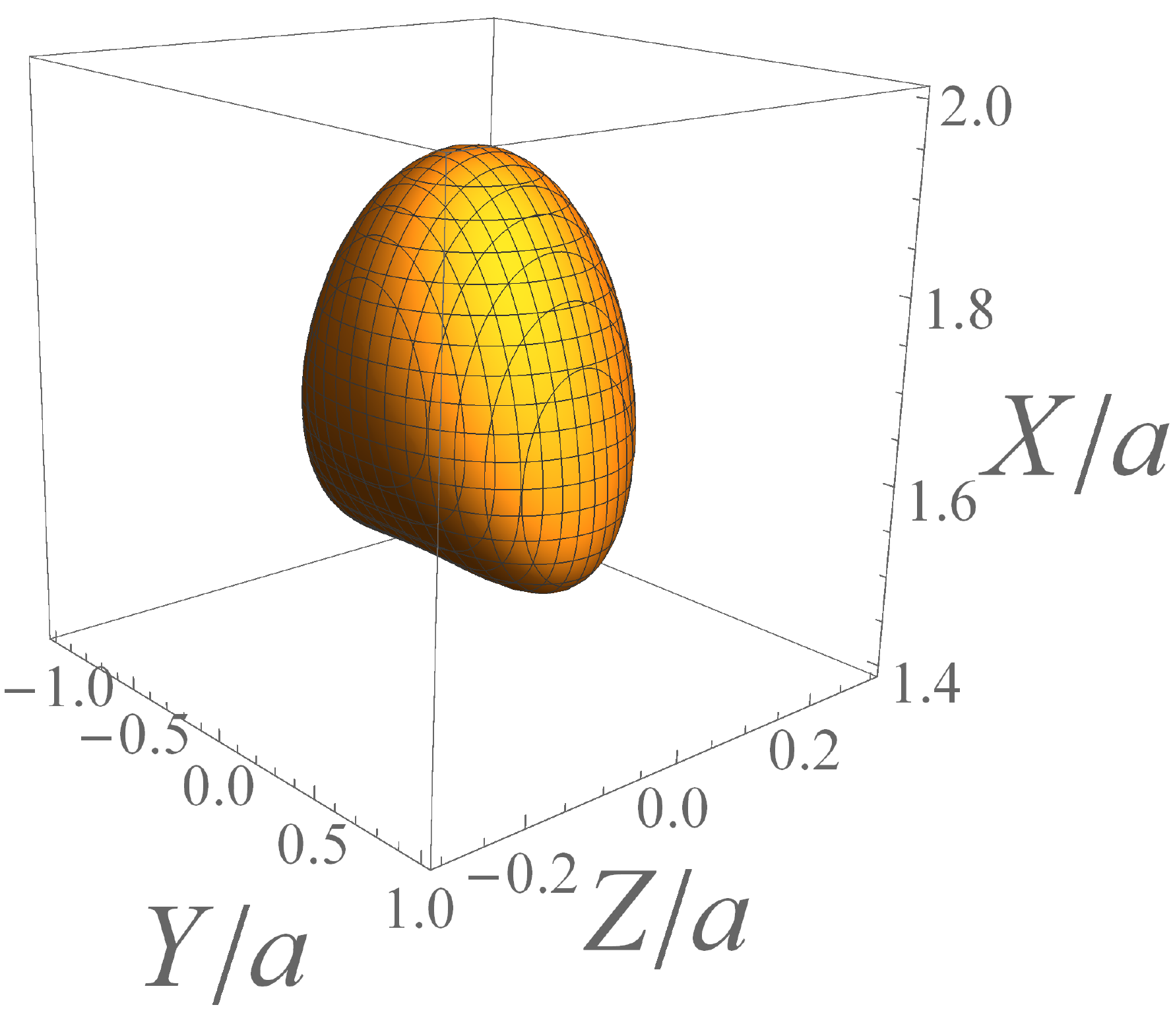}

     \end{subfigure}
     \\
     \vspace{1cm}
     \centering
     \begin{subfigure}[b]{\textwidth}
         \centering
         
         \caption{}
         
         \includegraphics[width=2.5cm]{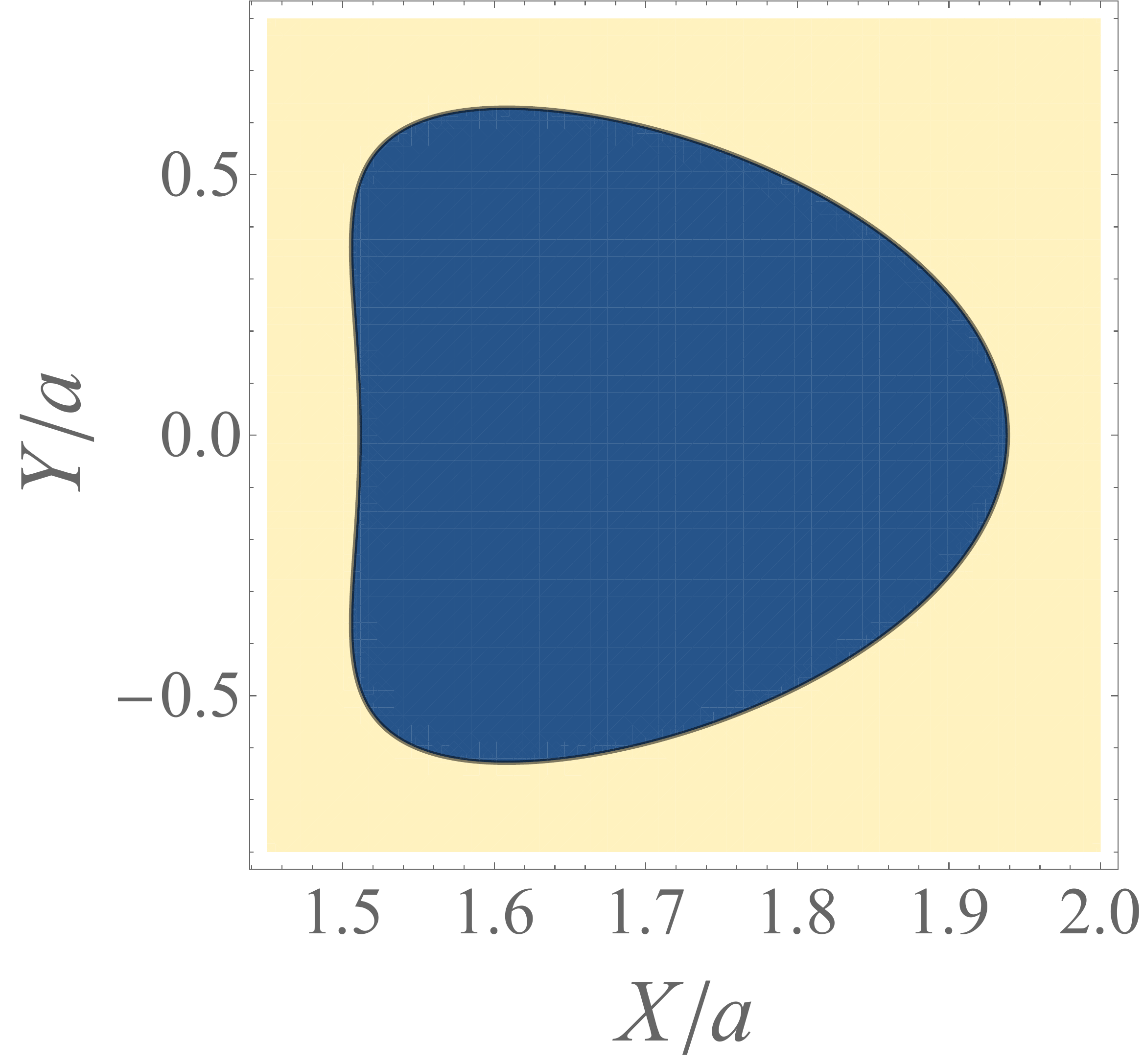}
         \hfill
         \includegraphics[width=2.5cm]{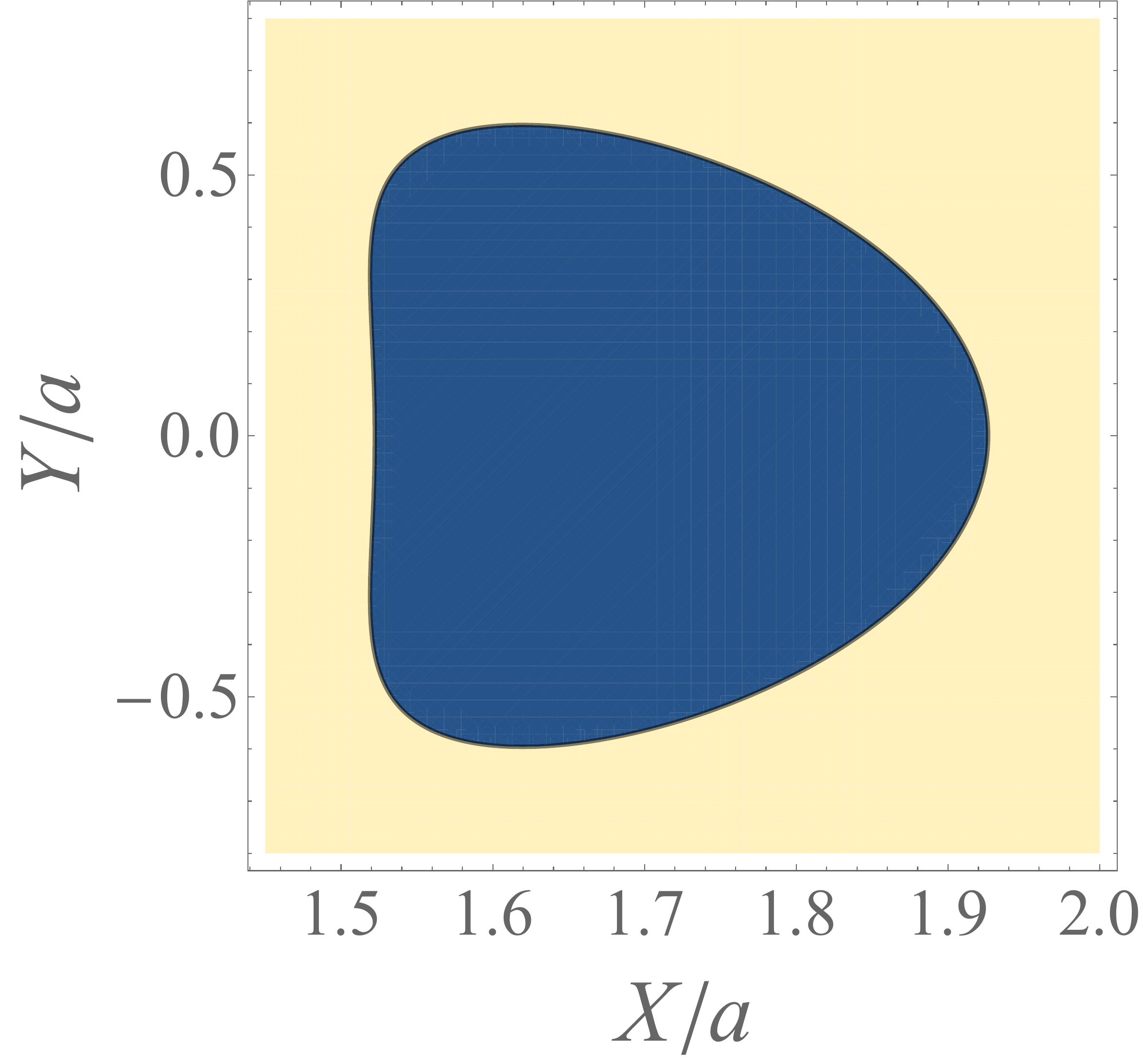}
         \hfill
         \includegraphics[width=2.5cm]{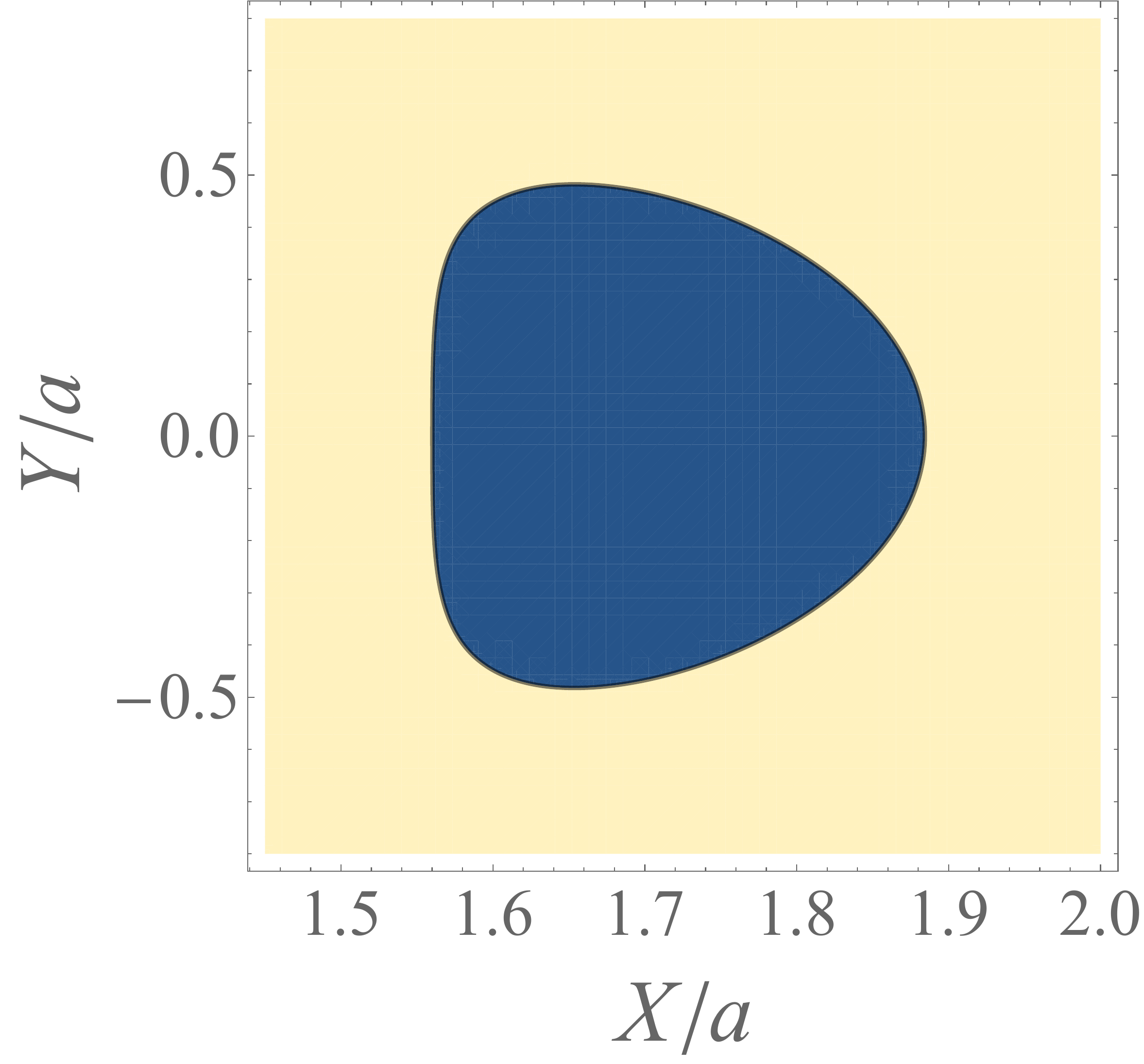}
         \hfill
         \includegraphics[width=2.5cm]{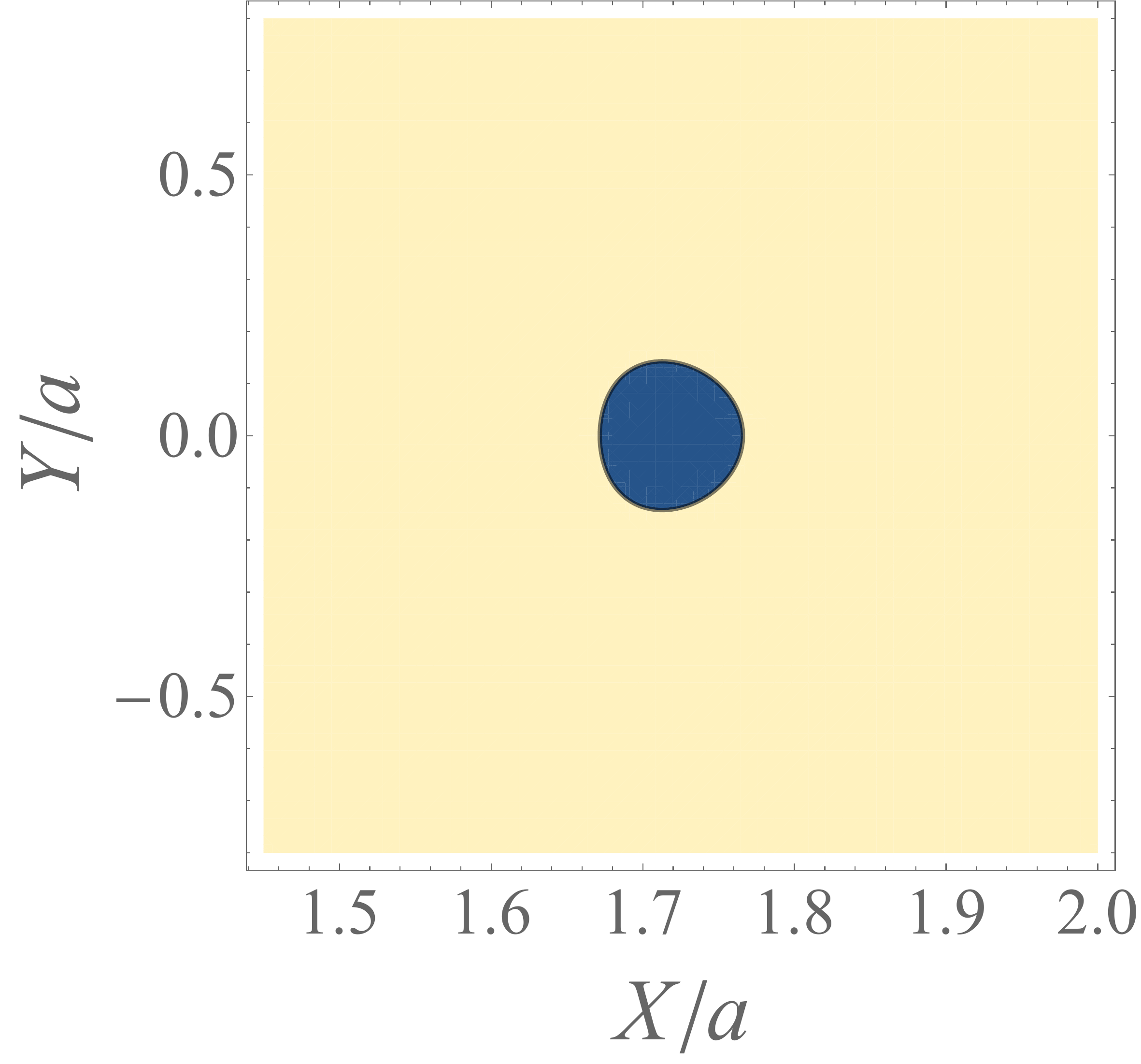}
         \hfill
         \includegraphics[width=2.5cm]{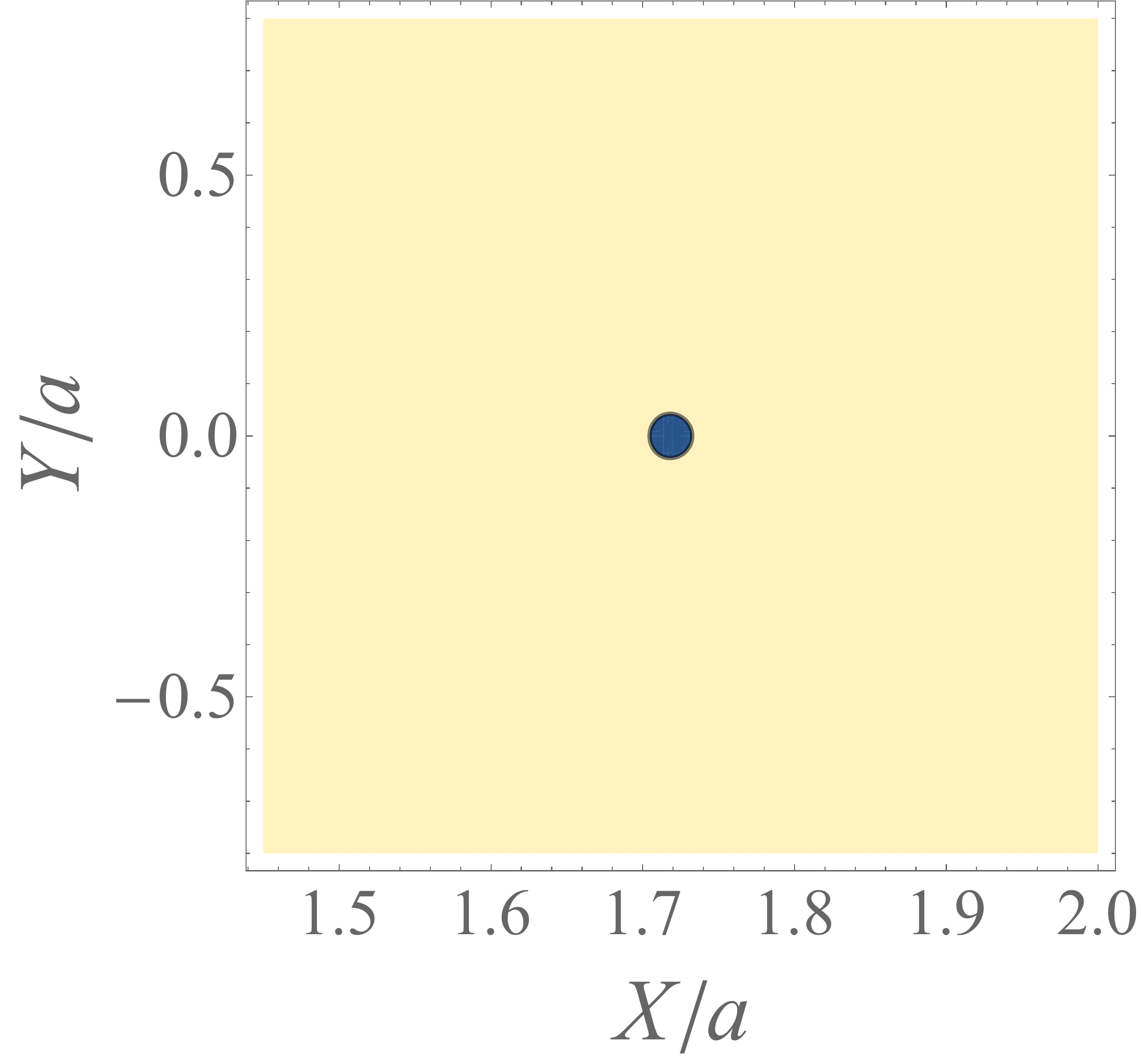}
         
     \end{subfigure}
        \caption{
        Trap obtained in Hund's case (b). (a) Three-dimensional plots of the two lobes (left) and upper lobe (right) of the trap located around $Z=0$. (b) Two-dimensional cuts of the upper lobe in the planes (from left to right) 
        $Z/a=0,0.05,0.1,0.15,0.1535$. The trap boundary is arbitrarily fixed at 
        $U=-3.8$ mK and one lobe has approximate extensions 
        $0.4a=80$ nm along $X$, 
        $1.3 a=260$ nm along $Y$ and 
        $0.3 a = 60$ nm along $Z$. The potential accommodates for an array of identical two-lobe traps periodic along $Z$ with period $\pi / \beta_2 \approx 530$ nm. }
\label{plot3D}       
\end{figure}

Approximating the potential around its minimum by three independent harmonic oscillators along $X$, $Y$ and $Z$ axes, \emph{i.e.} $U\left( X,Y,Z \right) \approx U_{\mbox{min}} + \frac{1}{2} \left( k_X X^2 + k_Y Y^2 + k_Z Z^2 \right)$, we numerically find the following values for the spring constants  $k_{j=X,Y,Z}$ and associated energies $\hbar \omega_j \equiv \hbar \sqrt{\frac{k_j}{m}}$
\begin{eqnarray*}
k_{X}\approx9.5\;\mbox{mK}\cdot a^{-2} & \qquad & \hbar \omega_{X}\approx26 \; \mu\mbox{K}\\
k_{Y}\approx1.0\;\mbox{mK}\cdot a^{-2} & \qquad & \hbar \omega_{Y}\approx9\; \mu\mbox{K}\\
k_{Z}\approx17.5\;\mbox{mK}\cdot a^{-2} & \qquad & \hbar \omega_{Z}\approx35 \; \mu\mbox{K}
\end{eqnarray*}
Figure \ref{plot3D} shows 3D views (subfigure a) and contour plots (subfigure b) of the two 
lobes of the trapping potential around $Z=0$. There, the boundary of the trap was arbitrarily fixed at $-3.8$ mK, which corresponds to a trap depth of $200$ nK. For this choice,  tunnel effect between different lobes was numerically checked to be negligible and each lobe can accomodate for about $500$ 
translational bound states of the molecule. The dimensions of the trap along $X$, $Y$ and $Z$ are respectively found to be approximately $0.4a=80$ nm, $1.3 a=260$ nm and $0.3 a = 60$ nm. 

Hund's case (a) leads to qualitatively similar results as case (b). Now, as shown in Table \ref{TabPolar}, the molecular polarisability tensor takes three different values corresponding to three groups of $\left( M,\Sigma \right)$ components, each group being associated to a specific alignement parameter $a_Z$. For each of these state manifolds, the general shape of the trap is the same as for Hund's case (b) but the position of the minimum differs in $R$, as can be seen in figures (\ref{contourplots},\ref{Potentialrhophiz}). 
The shape of the optical trap experienced by the molecule in its different states results from the combined effects of its alignment properties and the anisotropy of the electric field. 
The potential can indeed be written $U=\mathcal{V}+a_{Z}\mathcal{W}$,
with 
\begin{eqnarray*}
\mathcal{V} & \equiv\frac{1}{2}\sum_{j=1,2}\left\{ \left(\alpha_{\parallel}^{\left(j\right)}-\alpha_{\perp}^{\left(j\right)}\right)\left|E_{j,Z}^{\left(+\right)}\right|^{2}-\left(\alpha_{\parallel}^{\left(j\right)}+\alpha_{\perp}^{\left(j\right)}\right)\left|{\bf E}_{j}^{\left(+\right)}\right|^{2}\right\} \\
\mathcal{W} & \equiv\frac{1}{2}\sum_{j=1,2}\left(\alpha_{\parallel}^{\left(j\right)}-\alpha_{\perp}^{\left(j\right)}\right)\left(\left|{\bf E}_{j}^{\left(+\right)}\right|^{2}-3\left|E_{j,Z}^{\left(+\right)}\right|^{2}\right)
\end{eqnarray*}
where $E_{j,Z}^{\left(+\right)}$ denotes the $Z$ component of the
field $\mathbf{E}_{j}^{\left(+\right)}$. 
When $\mathcal{W}>0$ ($<0$), the potential increases (decreases)
with $a_{Z}$. This observation allows us to explain the ordering of
the curves in Fig. \ref{Potentialrhophiz} (c). Around the trap minimum $Z=0$, $R = R^{(s)}_{\mbox{min}}$
and $\Theta = 0$, $\mathcal{W}$ is found to be positive from inspection of the electric fields
and $U$ increases with $a_{Z}$ : the smallest trap minimum is hence
obtained for Hund's case (a) $\left(M,\Sigma\right)=\left(0,\pm1\right),\left(\pm1,0\right)$
characterized by $a_{Z}=0.2$, while the largest one is observed for
Hund's case (a) $\left(M,\Sigma\right)=\left(0,0\right)$ characterized
by $a_{Z}=0.6$. By contrast, around $Z=\frac{\pi}{2\beta_{2}}$,
$R = R^{(s)}_{\mbox{min}}$ and $\Theta = 0$, $\mathcal{W}$ is
negative and $U$ decreases with $a_{Z}$ : the order of the curves
is therefore inverted with respect to the previous case.

Since for $j=1,2$, $\alpha_{\parallel}^{\left(j\right)}>\alpha_{\perp}^{\left(j\right)}$, the term $\mathcal{W}$ directly reflects the anisotropy of the trapping fields. Note that, for our specific choice of \textsc{laser} frequencies and amplitudes, the term $\mathcal{W}\left(R,\Theta,Z\right)$ is dominated
by the standing-wave field contribution for $R\approx R^{(s)}_{\mbox{min}}$,
$\Theta\approx0$, $Z\approx0$ or $\frac{\pi}{2\beta_{2}}$. The
form of the standing-wave field in equation \ref{Estand} therefore defines the sign of $\mathcal{W}$.
For $Z=0$ the standing-wave
field is indeed purely transverse, hence 
$\mathcal{W}\left(R^{(s)}_{\mbox{min}},\Theta=0,Z=0\right)\approx\frac{1}{2}\left(\alpha_{\parallel}^{\left(2\right)}-\alpha_{\perp}^{\left(2\right)}\right)\left|{\bf E}_{2}^{\left(+\right)}\right|^{2}>0$.
Conversely, it is purely longitudinal for $Z=\frac{\pi}{2\beta_{2}}$
and hence $\mathcal{W}\left(R^{(s)}_{\mbox{min}},\Theta=0,Z=\frac{\pi}{2\beta_{2}}\right)\equiv-\frac{3}{2}\sum_{j=1,2}\left(\alpha_{\parallel}^{\left(j\right)}-\alpha_{\perp}^{\left(j\right)}\right)\left|E_{j,Z}^{\left(+\right)}\right|^{2}<0$.
To put it in a nutshell, aligning the molecule axis with the strong field axes favours trapping. 

\subsection*{Discussion}
We finish this section with a few remarks. First, we underline that, in the analysis above, we did not take into account the Casimir-Polder interaction between the molecule and the nanofibre.  We can give a rough estimate of the associated energy shift through approximating the fibre by a half dielectric space -- this crude approximation is all the better as the molecule is closer to the fibre and overestimates the Casimir-Polder shift induced by the presence of the fibre. Denoting by $D$ the distance of the molecule centre of mass from the surface of the half-space (we assume the X axis is orthogonal the medium surface), one gets the following expression for the Casimir-Polder shift $\delta E_n$ of the state $\left|\phi^s_n \right\rangle$ in the nonretarded regime \cite{EW03} 
\begin{eqnarray} 
\label{shift}
\delta E_n=&&-\frac{1}{4\pi\epsilon_0}\frac{1}{16 D^3} \frac{n_1^2-1}{n_1^2+1} \sum_{n'\neq n} \\
&&\left( \left|\langle \phi_n^s|d^s_Y|\phi_{n'}^s\rangle\right|^2 + \left|\langle \phi_n^s|d^s_Z|\phi_{n'}^s\rangle\right|^2  + 2 \left|\langle \phi_n^s|d^s_X|\phi_{n'}^s\rangle\right|^2 \right)
\nonumber
\end{eqnarray}
where $d^s_i$ ($i=X,Y,Z$) denotes the space fixed transition dipole vector (\ref{AppenPolarizability}). With $\left|\langle \phi_n^s|d^s_i|\phi_{n'}^s\rangle\right| \approx 4$ au  (see figure A1 in \cite{DDD15} and figure 7 in \cite{AA12}) and $D\approx a = 200$ nm, one finds $\delta E_n\approx 6 \; \mu\mbox{K}$ which is completely negligible with respect to \textsc{laser}-induced trapping potential.

We also want to emphasize that, in Hund's case (b), the molecule prepared in any of the states with magnetic numbers $M=0,\pm 1$ will be submitted to exactly the same trapping potential. Translational motion of the molecule in the trap will therefore cause no dephasing between the different $M$ components. A qutrit of information can hence be safely encoded on those states. In the same way, in case (a), two manifolds, each of which comprises four states, can be used to safely store two qubits of information. This is promising for future quantum technology uses of free molecule-nanofibre interfaces.

\section{Conclusion \label{Conclusion}}

This article presented a theoretical proposal of a two-colour optical trap for a diatomic molecule, Rb$_2$, prepared in the metastable state $(1)^3\Sigma^+_u$ implemented in the fundamental guided mode of a silica optical nanofibre. The envisioned setup was described in detail, including trapping \textsc{laser} beam frequencies, amplitudes and polarizations as well as molecular tensor polarisability. Different Hund's cases were investigated and the influence on trapping efficiency of alignment of the molecule with respect to the nanofibre axis was analyzed. 

Combining the richness of molecular state space with the potentialities of nanofibre-based setups, including chiral quantum optics, is very promising for, e.g., quantum simulation. The present article is a very preliminary step towards achieving such a molecule-nanofiber platform. 

Future works shall be devoted to the more detailed investigation of effects we have neglected or dismissed here, for sake of simplicity, such as the hyperfine structure of Rubidium \cite{STL10}. Other molecular species shall also be considered as well as the interactions between two molecules trapped in neighbouring sites.

\appendix

\section{Fundamental guided mode of the fibre} \label{AppGuidedMode}

In this appendix we recall the expression of quasi-linearly polarized
electric field of the fundamental guided mode HE$_{11}$ of an optical
nanofibre of radius $a$ and optical index $n_{1}$. We refer to figure
\ref{FigSystem} for the definition of Cartesian and cylindrical coordinate frames. 

The electric field at frequency $\omega$ decomposes into its positive-
and negative-frequency parts, respectively denoted by $\underline{{\bf E}}^{\left(+\right)}e^{-\mbox{i}\omega t}$
and $\underline{{\bf E}}^{\left(-\right)}e^{\mbox{i}\omega t}$, i.e.
${\bf E}=\underline{{\bf E}}^{\left(+\right)}e^{-\mbox{i}\omega t}+\underline{{\bf E}}^{\left(-\right)}e^{\mbox{i}\omega t}$,
with $\underline{{\bf E}}^{\left(-\right)}\equiv\left[\underline{{\bf E}}^{\left(+\right)}\right]^{*}$.
The positive-frequency component at point $M\left(R,\Theta,Z\right)$
of the field quasi-linearly polarized along $X$ takes the following
form (for $R > a$), expressed in the Cartesian frame, 
\begin{eqnarray*}
{\bf E}_{\mbox{HE}_{11}^{\left(X\right)}}^{\left(+\right)}\left(R,\Theta,Z\right) & = & \mbox{i}\mathcal{A}e^{\mbox{i}f\beta Z}\left(\frac{hJ_{1}\left(ha\right)}{qK_{1}\left(qa\right)}\right) \\
& \times & \left(\begin{array}{c}
\left(1-s\right)K_{0}\left(q R \right)+\left(1+s\right)K_{2}\left(q R \right)\cos2\Theta\\
\left(1+s\right)K_{2}\left(q R \right)\sin2\Theta\\
-\mbox{i}f\frac{2q}{\beta}K_{1}\left(q R \right)\cos\Theta
\end{array}\right)_{XYZ}
\end{eqnarray*}
where $f=\pm$ stands for the propagation direction, $\mathcal{A}$
is a real amplitude, $\left(J_{l},K_{l}\right)$ denote the $l^{\mbox{th}}$
Bessel function of the first kind and $l^{\mbox{th}}$ modified Bessel
function of the second kind, respectively, and
\begin{eqnarray}
h & \equiv\sqrt{k_{0}^{2}n_{1}^{2}-\beta^{2}},\quad q\equiv\sqrt{\beta^{2}-k_{0}^{2}n_{2}^{2}},\quad k_{0}\equiv\frac{\omega}{c} \label{eqhq}\\
s & \equiv\left[\frac{1}{\left(ha\right)^{2}}+\frac{1}{\left(qa\right)^{2}}\right]\left[\frac{J_{1}'\left(ha\right)}{haJ_{1}\left(ha\right)}+\frac{K_{1}'\left(qa\right)}{qaK_{1}\left(qa\right)}\right] \label{eqs}
\end{eqnarray}
Finally, $\beta$ is known as the propagation constant and the solution
of the eigenvalue equation \cite{Mar89,SL83}
\begin{eqnarray}
\frac{J_{0}\left(ha\right)}{haJ_{1}\left(ha\right)} & =-\frac{n_{1}^{2}+n_{2}^{2}}{2n_{1}^{2}}\frac{K_{1}'\left(qa\right)}{qaK_{1}\left(qa\right)}+\frac{1}{\left(ha\right)^{2}} \label{eveq}\\
 & -\sqrt{\left[\frac{n_{1}^{2}-n_{2}^{2}}{2n_{1}^{2}}\frac{K_{1}'\left(qa\right)}{qaK_{1}\left(qa\right)}\right]^{2}+\left(\frac{\beta}{n_{1}k}\right)^{2}\left[\frac{1}{\left(qa\right)^{2}}+\frac{1}{\left(ha\right)^{2}}\right]^{2}} \nonumber
\end{eqnarray}

\begin{figure*}
\begin{centering}
\includegraphics[width=14cm]{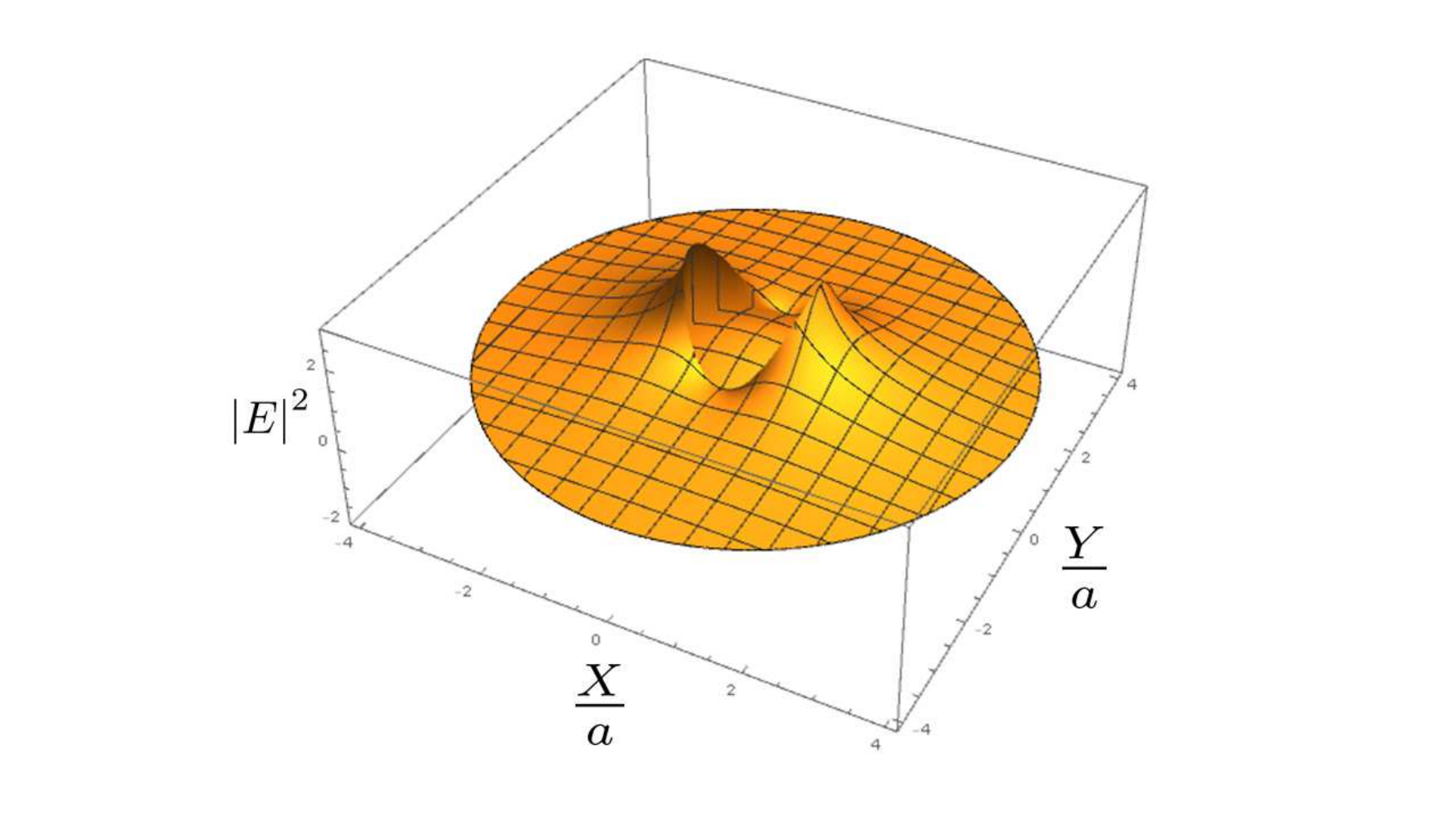}
\end{centering}
\caption{Intensity distribution in a transverse plane of the fundamental guided mode HE$_{11}^{(X)}$ quasi-linearly polarized along $(OX)$.}\label{GuidedModeIntensity}
\end{figure*}

The amplitude $\mathcal{A}$ can be related to the the power of the
\textsc{laser} beam, $\Pi$, as follows
\[
\Pi=\mathcal{A}^{2}\times\frac{4\pi a^{2}}{\mu_{0}c}\frac{\beta}{k}\times\left\{ \begin{array}{c}
\left(1+s^{2}+\frac{h^{2}}{\beta^{2}}\right)\left[J_{0}^{2}\left(ha\right)+J_{1}^{2}\left(ha\right)\right]\\
-\frac{2}{\left(ha\right)^{2}}\left(1+s\right)\left(1+s+\frac{h^{2}}{\beta^{2}}\right)J_{1}^{2}\left(ha\right)\\
+\left(\frac{hJ_{1}\left(ha\right)}{qK_{1}\left(qa\right)}\right)^{2}\left[\begin{array}{c}
\left(1+s^{2}-\frac{q^{2}}{\beta^{2}}\right)\left[K_{1}^{2}\left(qa\right)-K_{0}^{2}\left(qa\right)\right]\\
+\frac{2}{\left(qa\right)^{2}}\left(1+s\right)\left(1+s-\frac{q^{2}}{\beta^{2}}\right)K_{1}^{2}\left(qa\right)
\end{array}\right]
\end{array}\right\} 
\]
The intensity distribution of this field in a transverse plane is plotted in figure \ref{GuidedModeIntensity}.

When two counter-propagating fields travel in the fundamental guided
mode quasi-linearly polarized along $\left(OX\right)$ they create
a standing-wave whose electric field positive-frequency component
writes
\begin{eqnarray}
{\bf E}_{\mbox{HE}_{11}^{\left(X\right)},\mbox{sw}}^{\left(+\right)}\left(R,\Theta,Z\right) & = & 2\mbox{i}\mathcal{A}\left(\frac{hJ_{1}\left(ha\right)}{qK_{1}\left(qa\right)}\right) \label{Estand}\\
& \times &
\left(\begin{array}{c}
\left[\left(1-s\right)K_{0}\left(q R \right)+\left(1+s\right)K_{2}\left(q R \right)\cos2\Theta\right]\cos\left(\beta Z\right)  \\
\left(1+s\right)K_{2}\left(q R \right)\sin2\Theta\cos\left(\beta Z\right)\\
\frac{2q}{\beta}K_{1}\left(q R \right)\cos\Theta\sin\left(\beta Z\right)
\end{array}\right)_{XYZ} \nonumber
\end{eqnarray}

\section{Molecular polarisability tensor} \label{AppenPolarizability}
The dynamic polarisability at frequency $\omega$ of an atomic or a molecular system 
prepared in a quantum state  $ | \phi_n^s \rangle$ is
generally a tensor $\overline{\overline{\boldsymbol{\alpha}}}$ whose spherical components are given by
(equation (2.2) of ref. \cite{BBU22}):
\begin{equation} 
\alpha_{\mu \nu}(\phi_n^s, \omega)=\sum_{n'} \frac{2\omega_{n'n}}{\hbar(\omega_{n'n}^2-\omega^2)}
\langle \phi_n^s|d^s_\nu|\phi_{n'}^s\rangle \langle \phi_{n'}^s|d^s_\mu|\phi_n^s \rangle
\label{polar}
\end{equation}
where $\mu,\,\nu=-1,0,1$ label spherical components
, $n$ and $n'$ refer to sets of quantum numbers characterising 
the initial $|\phi_n^s\rangle$ and final $|\phi_{n'}^s\rangle$  states of the transitions at frequency $\omega_{n'n}$. Whereas these transitions involve only electronic excitations in atomic systems, they involve vibrational and rotational ones as well in the diatomic molecular case which we focus on in this appendix. 

The superscript $s$ 
emphasizes that the initial and final states as well as the spherical components of the transition dipole operator $d^s_\mu$ are defined with respect to the space-fixed frame (figure \ref{FigSystem}). Molecular data, however, like transition dipoles, are known in the body-fixed frame. It is therefore necessary to perform a transformation between space-fixed and body-fixed frames to calculate the polarisability tensor. The space-fixed components can be expressed in terms of the body-fixed ones using \cite{Zare88}, equation (3.103):
\begin{equation}
d^s_\mu=\sum_{m=-1,0,+1} D^{1\;*}_{\mu \, m} (\phi,\theta,\chi) d^b_m  
\label{dsphe}
\end{equation}
where the superscript $b$ refers to the body-fixed components (figure \ref{FigSystem}).  
Similarly, space-fixed molecular wavefunctions must be expressed in terms of body-fixed ones.
This task is easier if we can approximate the molecular states by Hund's limiting cases (ref. \cite{Brion04}, p. 100). Below we consider Hund's cases (a) and (b), which are the most common.  

Hund's case (a) wavefunctions are labelled by the set of quantum numbers $n=(\Lambda, S, \Sigma, v, J, \Omega, M)$ (ref. \cite{Zare88}, p. 298).
In this case, electrostatic interaction is strong enough so that the $z$ component $\Lambda$ of the electronic orbital angular momentum is an approximate good quantum number. Similarly, the spin-orbit coupling is strong enough in this case so that the $z$ component $\Sigma$ of the electronic spin $S$ is also an approximate good quantum number. $v$ is the vibrational quantum number, $J$ the total angular momentum, $\Omega=\Lambda+\Sigma$ and $M$ its components on the body-fixed $z$ and space-fixed $Z$  axes, respectively.
The space-fixed molecular wavefunction can thus be written: 
\begin{equation}
\label{phimol}
|\phi_n^s\rangle=|\phi^b_n\rangle |J \Omega M \rangle
\label{phisa}
\end{equation}
where  $|J \Omega M \rangle$ is a symmetric top rotational wavefunction (see \cite{Zare88}, equation (3.125), p. 105) and $|\phi^b_n \rangle$ the body-fixed molecular one. This in turn can be decomposed into orbital, spin and vibrational components (see \cite{Zare88}, equation (2) p. 298):
\begin{equation}
\label{phiba}
|\phi^b_n\rangle=|\Lambda v J \rangle |S\Sigma \rangle
\end{equation} 
where $|\Lambda v J \rangle $ is the product of the electronic state $\Lambda$ and of the vibrational state $v$ supported by the electronic potential associated to $\Lambda$. It may (slightly) depend on the rotational state $J$ of the molecule.    
Then inserting equations (\ref{dsphe}, \ref{phisa}, \ref{phiba}) into equation (\ref{polar}), we obtain
\begin{eqnarray}
\label{phiphi}
 &\langle \phi_n^s|d^s_\nu|\phi_{n'}^s\rangle  \langle \phi_{n'}^s|d^s_\mu|\phi_n^s \rangle = 
\delta_{\nu\,-\mu} \, \delta_{S'S} \, \delta_{\Sigma'\Sigma} \, \delta_{M' M+\mu} \\ & \times (-1)^{\mu} (2J+1)(2J'+1) \times |\langle \Lambda v J| d^b_{\Lambda-\Lambda'} | \Lambda' v' J' \rangle|^2 \nonumber \\  
& \times \left( \begin{array}{ccc}
J & 1 & J' \\
-M &-\mu & M'
\end{array}\right)^2
\left( \begin{array}{ccc}
J & 1 & J' \\
-\Lambda-\Sigma &\Lambda-\Lambda' & \Lambda'+\Sigma'
\end{array} \right)^2
\nonumber 
\end{eqnarray}
where $n'$ denotes the set of quantum numbers $(\Lambda', S', \Sigma', v', J', \Omega', M')$ which label the excited states of the transitions appearing in equation (\ref{polar}). 
The 3-j coefficients are known analytically (ref. \cite{Zare88}, equation (2.25) p. 49) and the body-fixed transition dipole matrix elements $\langle \Lambda v J| d^b_{\Lambda-\Lambda'} | \Lambda' v' J' \rangle$ are known from quantum  chemistry calculations.

If rotational interaction is significant, Hund's case (b) is more appropriate. In this approximation scheme, $\Sigma$ is no longer a good quantum number. By contrast the norm of the total angular momentum without spin, $\mathbf{N}=\mathbf{J}-\mathbf{S}$, becomes a good quantum number, and so is its projection onto the body-fixed frame $z$ axis, $\Lambda$, as it was in Hund's case (a) (ref. \cite{Brion04}, p. 103).  Finally, in Hund's case (b) the set of good quantum numbers is :
$n=(\Lambda, S, N, v, J , M)$ and the corrresponding basis is obtained by recoupling the states $| J \Omega M \rangle$ and $|S \Sigma \rangle$
(ref. \cite{Brion04}, 3.2.4b p. 103 and equation (3.2.61) p. 130)~:
\begin{equation}
\label{hundb}
|JMSN\Lambda\rangle = \sum_{\Sigma=-S}^{+S} \langle J \Omega, S \, -\Sigma=\Lambda-\Omega| J S N \Lambda \rangle  |J \Omega M \rangle | S \Sigma \rangle  
\end{equation}
which involves a Clebsh-Gordan coefficient. 
This equation can be used with equation (\ref{polar}) to compute Hund's case (b) polarisability and we obtain :
\begin{eqnarray}
\label{hundbf2}
&\langle \phi_n^s|d^s_\nu|\phi_{n'}^s \rangle \langle \phi_{n'}^s|d^s_\mu|\phi_n^s \rangle=\delta_{\mu \, -\nu} \delta_{S'S} \delta_{M'\,M+\mu}(-1)^{2M+\mu}
\\ & \times (2N+1)(2N'+1)(2J+1)(2J'+1)   |\langle \Lambda v J| d^b_{\Lambda-\Lambda'} | \Lambda' v' J' \rangle|^2 \nonumber \\
& \times
\left\{ \begin{array}{ccc}
1 & J' & J  \\
S & N & N'
\end{array} \right\}^2
\left( \begin{array}{ccc}
N & 1 & N' \\
-\Lambda &\Lambda-\Lambda' & \Lambda'
\end{array} \right)^2
\left( \begin{array}{ccc}
J & 1 & J' \\
-M &-\mu & M'
\end{array}\right)^2
\nonumber 
\end{eqnarray}
where the quantity in curly brackets is a 6-j coefficient (ref. \cite{Zare88}, equation (4.8) p. 145).

\ack{This research has been partially funded by l’Agence Nationale de la Recherche (ANR), project ANR-22-CE47-0011, and through the EUR grant NanoX n° ANR-17-EURE-0009 in the framework of the "Programme des Investissements d’Avenir". For the purpose of open access, the authors have applied a CC-BY public copyright licence to any Author Accepted Manuscript (AAM) version arising from this submission.  E. B. dedicates this work to the memory of V. M. Akulin.}

\end{document}